\documentclass[%
 reprint,
amsmath,amssymb,
pra,
]{revtex4-1}
\providecommand{\c}{\mathrm{c}}
\usepackage[english]{babel}
\usepackage{threeparttable}
\usepackage{caption}
\usepackage{caption}
\usepackage{mathtools}

\DeclarePairedDelimiter\floor{\lfloor}{\rfloor}
\usepackage{mathdots}
\usepackage[utf8]{inputenc}
\usepackage{times}
\usepackage{newtxtext}
\usepackage{mathptmx}
\usepackage{amsthm}
        \usepackage[top=30mm, bottom=30mm, left=25mm, right=25mm]{geometry}
 \usepackage{amsmath}
        \usepackage{amsmath, amssymb}
	\usepackage{color}
	\usepackage{graphicx}
	\usepackage{braket}
	\usepackage[subrefformat=parens]{subcaption}
	\usepackage{txfonts}
	\usepackage{enumerate}
	\usepackage{bm}

	\definecolor{gray}{gray}{0.5}
	\definecolor{emerald}{cmyk}{1,0,0.5,0}
	\definecolor{bluegreen}{cmyk}{0.85,0,0.33,0}
	\definecolor{violet}{cmyk}{0.79,0.88,0,0}

\newcommand{\1}{\mbox{1}\hspace{-0.23em}\mbox{l}}

\begin{document}
\preprint{APS/123-QED}

\title{Dissipative time crystals originating from parity-time symmetry}

\author{Yuma Nakanishi}
 \altaffiliation
 {}
  \email{nakanishi.y@stat.phys.titech.ac.jp}
\author{Tomohiro Sasamoto}%
\affiliation{%
Department of Physics, Tokyo Institute of Technology, 2-12-1 Ookayama
Meguro-ku, Tokyo, 152-8551, JAPAN}%

\date{\today}

\begin{abstract}
This study aims to provide evidence regarding the emergence of a class of dissipative time crystals when $\mathcal{PT}$
symmetry of the systems is restored in collective spin systems with Lindblad dynamics. First, we show that a standard model of boundary time crystals (BTCs) satisfies the Liouvillian $\mathcal{PT}$ symmetry, and prove that BTC exists only when the stationary state is $\mathcal{PT}$ symmetric in the large-spin limit. Also, a similar statement is confirmed numerically for another BTC model. In addition, the mechanism of the appearance of BTCs is discussed through the development of a perturbation theory for a class of the one-spin models under weak dissipations. Consequently, we show that BTCs appear in the first-order correction when the total gain and loss are balanced. These results strongly suggest that BTCs are time crystals originating from $\mathcal{PT}$ symmetry.
\end{abstract}

\maketitle

\textit{Introduction}. Crystals are ubiquitous many-body systems wherein continuous \textit{space}-translation symmetry is spontaneously broken. Similarly, dynamic many-body states that spontaneously break continuous \textit{time}-translation symmetry, namely (continuous) time crystals, were proposed by Wilczek in 2012 $\text{\cite{Wilczek}}$. However, it has been proven that time crystals do not exist in ground and equilibrium states, at least for long-range interacting systems $\text{\cite{Watanabe,Watanabe2}}$. In non-equilibrium systems such as Floquet systems $\text{\cite{Else,Zhang,Russomanno,Yao,Huang,Sacha1}}$ and dissipative systems $\text{\cite{Iemini,Piccitto,dos,Minganti4,Booker,Tucker,Lled2,Lled,Federico,Michal,AC,Buca1,Ke,BB,Sacha2}}$, (discrete or dissipative) time crystals have been observed theoretically and experimentally. 

Dissipative time crystals are non-trivial states characterized by persistent periodic oscillations at late times induced by coupling with the external environment $\text{\cite{Booker}}$. In particular, a kind of dissipative time crystal, called boundary time crystal (BTC), has been often studied recently $\text{\cite{Iemini,Piccitto,dos,Minganti4,Federico,Michal,AC}}$. 
BTCs were first introduced using a collective spin model with Lindblad dynamics $\text{\cite{Iemini}}$, which describes a collection of spin-1/2 with all-to-all couplings interacting collectively with external Markovian baths. This model could be derived by tracing out the bulk (environment) degrees of freedom while leaving the boundary (system) degrees of freedom. Further, it has been confirmed that persistent oscillatory phenomena at late times emerge only \textit{in the thermodynamic limit.} Even though such phenomena were already noted 40 years earlier as cooperative resonance fluorescence $\text{\cite{Carmichael}}$, it should be emphasized that there are various novel aspects in recent studies of BTCs. In particular, the importance of Liouvillian eigenvalues has been realized $\text{\cite{Iemini,Piccitto,Minganti4}}$ because the dynamics can be fully understood in terms of their eigenvalues and eigenmodes. In addition, recent developments of the spectral theory of dissipative phase transitions $\text{\cite{Minganti2,Kessler}}$ and exact solutions of the Liouvillian spectrum $\text{\cite{Nakagawa,McDonald,bbuca,Prosen6,Pedro,Prosen3,Prosen4}}$ have also increased the interest in investigating Liouvillian eigenvalues.

The dynamical properties of BTCs are often investigated via numerical calculations of Liouvillian eigenvalues $\text{\cite{Iemini,Piccitto}}$, mean-field approximation method $\text{\cite{Iemini,Piccitto,dos,Carmichael}}$, and quantum trajectory method $\text{\cite{Link}}$. In particular, BTCs must satisfy two conditions for the Liouvillian spectrum, which characterize non-stationary periodic oscillations at late times: (i) there exist pure imaginary eigenvalues $i\lambda_{j}$ and (ii) the quotient of each pure imaginary eigenvalue is a rational number $\lambda_{j}/\lambda_{k}\in\mathbb{Q}$ for all $j,\ k$. Moreover, BTCs are characterized by static properties such as the existence of a highly mixed and low-entangled eigenmode with a zero eigenvalue $\text{\cite{Piccitto,AC,Hannukainen,Puri,Lawande}}$. Also, the necessity of Hamiltonians' $\mathbb{Z}_{2}$ symmetry has been argued recently $\text{\cite{Piccitto}}$. 
However, the physical origin of the emergence of BTCs has not yet been elucidated, and most studies on Liouvillian eigenvalues for BTCs have been numerical.

Phase transitions accompanied by parity-time ($\mathcal{PT}$) symmetry breaking, namely $\mathcal{PT}$ phase transitions $\text{\cite{BenderC.M.Boettcher,MostafazadehA1}}$, are also phenomena wherein persistent oscillations emerge at late times in non-equilibrium systems. These are well-known phenomena in the context of non-Hermitian Hamiltonians (NHHs) $\text{\cite{Ashidasan}}$ with exactly balanced gain and loss, and have been widely investigated in a variety of physical experimental systems, such as mechanics $\text{\cite{Bender}}$, photonics $\text{\cite{RoterC}}$, plasmonics $\text{\cite{Alaeian}}$, electronics $\text{\cite{Schindler}}$, and open quantum systems without quantum jumps $\text{\cite{Wu1}}$. Mathematically, the Hamiltonian $H$ is considered to be $\mathcal{PT}$ symmetric if it holds that $[H, PT]=0$, where $P$ is a parity operator and $T$ is a time reversal operator $\text{\cite{BenderC.M.Boettcher,MostafazadehA1}}$. 
In addition, $\mathcal{PT}$ phase transitions in systems with Lindblad dynamics, hereafter referred to as Liouvillian $\mathcal{PT}$ phase transitions, have been recently discussed, and their understanding has been progressing $\text{\cite{Nakanishi,Huber1,Huber2,Prosen1,Van,Prosen2,Huybrechts}}$. 

This study attempts to demonstrate the emergence of a class of dissipative time crystals when $\mathcal{PT}$ symmetry is realized. 
We focus on a specific class of systems, collective spin systems with Lindblad dynamics, and provide the results about the Liouvillian eigenvalues and stationary state for some specific examples. 
First, we show that the $\mathcal{PT}$ symmetric phase of an open two-spin model with Liouvillian $\mathcal{PT}$ symmetry $\text{\cite{Nakanishi,Huber1,Huber2}}$ is a BTC. Here, an $n$-spin model is a system with $n$-collective spin operators in the interaction term. Second, we show that an open collective spin model with interaction owing to a transverse magnetic field and excitation decay (hereafter referred to as the one-spin BTC model) satisfies the proposed definition of the Liouvillian $\mathcal{PT}$ symmetry $\text{\cite{Huber1,Huber2}}$ if the parity transformation is appropriately chosen. In addition, we prove that the $\mathcal{PT}$ symmetry breaking of the stationary state occurs at the BTC phase transition point in the large-spin limit. Next, we confirm that the generalized one-spin BTC model studied in Ref.$\text{\cite{Piccitto}}$ also has Liouvillian $\mathcal{PT}$ symmetry. Further, we numerically show that the stationary state exhibits $\mathcal{PT}$ symmetry in the BTC phase. Finally, we perform a perturbative analysis of a class of one-spin models, including the one-spin BTC model under weak dissipation. Consequently, we show that BTCs appear in the first-order correction owing to the balanced total gain and loss. These results strongly suggest that BTCs in collective spin systems are time crystals originating from $\mathcal{PT}$ symmetry. \\

\textit{Liouvillian spectrum and Liouvillian $\mathcal{PT}$ symmetry}. 
In open quantum systems where the evolution of states is completely positive and trace-preserving (CPTP) Markovian, the time evolution of the density matrix $\rho(t)$ is described by the Lindblad master equation (GKSL equation) $\text{\cite{Lindblad,Breuer,ARivas,GKS}}$ as follows
\begin{eqnarray}
\label{lindblad}
\frac{d\rho}{dt}=-i[H,\rho(t)]+\sum_{i}\mathcal{D}[L_{i}]\rho,
\end{eqnarray}
where $H$ is a Hamiltonian, $L_{i}$ is the Lindblad operator, and the dissipation superoperators $\mathcal{D}[L_{i}]$ are defined as $\mathcal{D}[L_{i}]\rho=2L_{i}\rho L_{i}^{\dagger}-L_{i}^{\dagger}L_{i}\rho-\rho L_{i}^{\dagger}L_{i}.$
Here, index $i$ labels the Lindblad operators. 

The Lindblad master equation ($\ref{lindblad}$) is linear in $\rho$, thus, it can be rewritten with a superoperator, which is a linear operator acting on a vector space of linear operators, as follows:
\begin{align}
\label{120}
\frac{d\rho(t)}{dt}=\hat{\mathcal{L}}\rho(t).
\end{align}
Here, $\hat{\mathcal{L}}$ is referred to as the Liouvillian
superoperator. 

The eigenvalues $\lambda_{i}$ and eigenmodes $\rho_{i}$ of the Liouvillian can be obtained by solving the following equation:
\begin{align}
\label{1.8}
\hat{\mathcal{L}}\rho_{i}=\lambda_{i}\rho_{i}.
\end{align}
It is generally known that Re[$\lambda_{i}$]$\leq0, \forall i$; if $\hat{\mathcal{L}}\rho_{i}=\lambda_{i}\rho_{i}$, then $\hat{\mathcal{L}}\rho_{i}^{\dagger}=\lambda_{i}^{*}\rho_{i}^{\dagger}$ $\text{\cite{Breuer,ARivas}}$. Here, we assume the existence of a unique steady state and set the eigenvalues as $0=|\textrm{Re}[\lambda_{0}]|<|\textrm{Re}[\lambda_{1}]|\leq|\textrm{Re}[\lambda_{2}]|\leq\cdots$. The steady state is then written as $\rho_{ss}=\rho_{0}/\textrm{Tr}[\rho_{0}] $. In addition, the absolute value of the real part of the second maximal eigenvalue is referred to as the Liouvillian gap $\text{\cite{Kessler,Minganti2}}$ and determines the slowest relaxation rate. Closing the Liouvillian gap is necessary for dissipative phase transitions in steady state $\text{\cite{Kessler,Minganti2}}$.

It should be noted that imaginary eigenvalues emerge only in the thermodynamic limit for Liouvillian $\mathcal{PT}$ phases and BTCs. Therefore, the thermodynamic limit and the long-time limit are not commutative, that is, $\lim_{S\to\infty}\lim_{t\to\infty}\rho(t)\neq\lim_{t\to\infty}\lim_{S\to\infty}\rho(t)$ $\text{\cite{Pei}}$. In the former case, the steady state is static without oscillation. We refer to the state $\lim_{t\to\infty}\rho(t)$ as the "\textit{stationary state}", whereas, in the latter case, the state at late times includes oscillating non-decay modes.

Many studies on Liouvillian $\mathcal{PT}$ symmetry have been conducted recently $\text{\cite{Nakanishi,Prosen1,Huber1,Huber2,Prosen2,Huybrechts,Van}}$; however, the definition of Liouvillian $\mathcal{PT}$ symmetry has not been uniquely determined yet. In our arguments, we adopted (a slightly modified version of ) the definition proposed in Ref.$\text{\cite{Huber2}}$ because similar properties to those of NHH $\mathcal{PT}$ phase transitions have been confirmed in a specific two-spin model that satisfies this definition, as mentioned below. A Liouvillian associated with the Lindblad equation (\ref{lindblad}) is considered to be $\mathcal{PT}$ symmetric if the following relation holds.
\begin{align}
\label{HuberPT}
\hat{\mathcal{L}}[\mathbb{PT}(H);\mathbb{PT}^\prime(L_\mu),\mu=1,2,\cdots]=\hat{\mathcal{L}}[H;L_\mu,\mu=1,2,\cdots],
\end{align}
where $\mathbb{PT}(H)=PTH(PT)^{-1}=P\bar{H}P^{-1}$, $\mathbb{PT}^\prime(L_{\mu})=PL_{\mu}^{\dagger}P^{-1}$. Here, $P$ and $T$ are parity and time reversal operators, and $\bar{H}$ means the complex conjugation of $H$. Further, $\mathbb{PT}(H)$ denotes the conventional $\mathcal{PT}$ transformation of Hamiltonian. However, in the $\mathbb{PT}^{\prime}$ transformation, the time-reversal transformation of dissipations represents an exchange of creation and annihilation operators. \\

\begin{figure}[t]
   \vspace*{-0.9cm}
     \hspace*{-5.4cm}
\includegraphics[bb=0mm 0mm 90mm 150mm,width=0.35\linewidth]{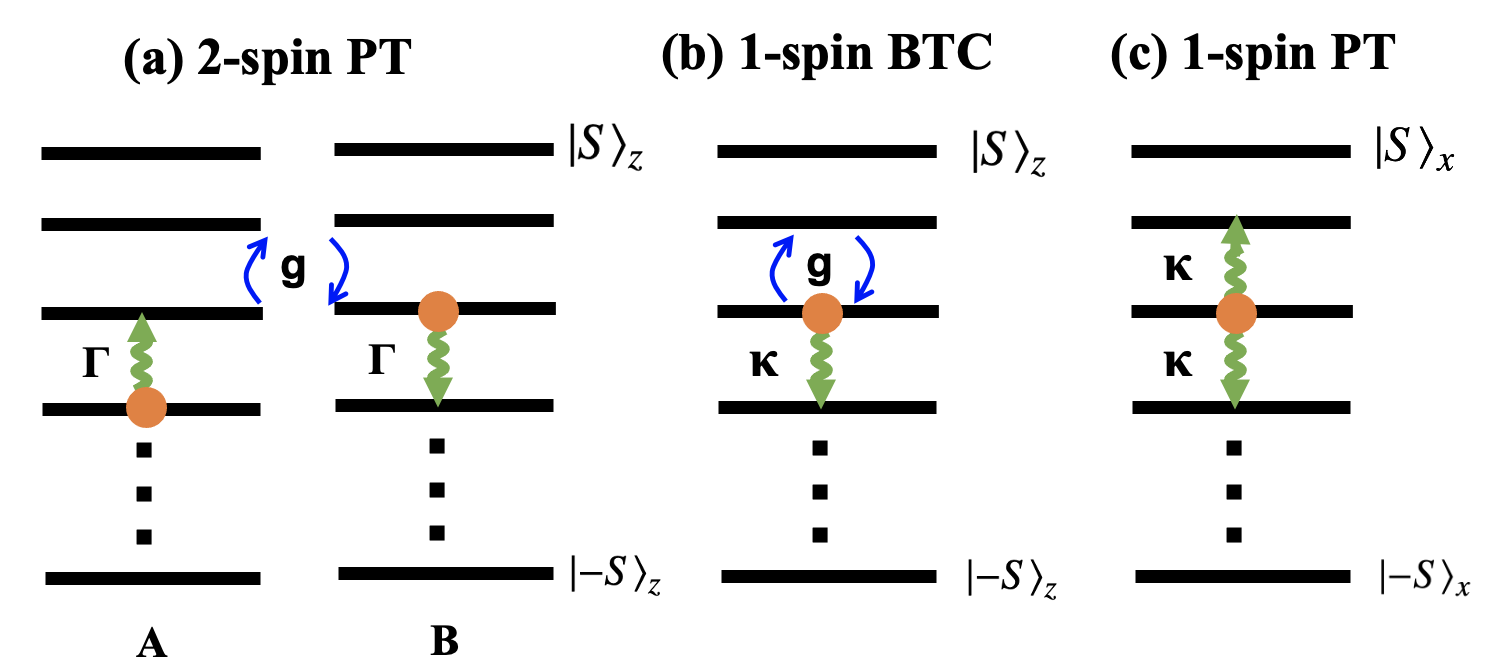}
    \caption{Illustration of dissipative spin-$S$ models, (a) two-spin $\mathcal{PT}$ model (b) one-spin BTC model (c) one-spin $\mathcal{PT}$ model. These models satisfy Liouvillian $\mathcal{PT}$ symmetry ($\ref{HuberPT}$) when the parity operator is (a) the exchange of two spins, (b) the reflection of the basis of $S_{z}$, (c) the identity operator or the reflection of the basis of $S_{z}$.}
    \label{figponti}
\end{figure}

\textit{Two-spin Liouvillian $\mathcal{PT}$ symmetric model is BTC}.
To investigate the usefulness of the definition ($\ref{HuberPT}$), an open two-spin-$S$ model with exactly balanced gain and loss [Fig.$\ref{figponti}$ (a)] was actively investigated $\text{\cite{Nakanishi,Huber1,Huber2}}$. Here, $S$ denotes the total spin. Denoting the subsystems as A and B, the Lindblad equation is expressed as
\begin{align}
\label{2spindefinition2}
\frac{\partial}{\partial t}\rho=-ig[H,\rho]+\frac{\Gamma}{2S}\mathcal{D}[S_{+,A}]\rho+\frac{\Gamma}{2S}\mathcal{D}[S_{-,B}]\rho,
\end{align}
where $H=(S_{+,A}S_{-,B}+ \textrm{H.c.})/2S$, $S_{\pm}=S_{x}\pm iS_{y}$, and $g$ and $\Gamma$ are the strengths of the interaction and dissipation, respectively. Note that it decays to a unique steady state for a finite $S$ as spin ladder operators $S_{\pm}$ exist as one of the dissipation operators $\text{\cite{Nigro}}$. 

This model satisfies the definition ($\ref{HuberPT}$) when the parity transformation is the exchange of two spins, and a dissipative phase transition occurs at $\Gamma=g$ when $S=\infty$. Moreover, the symmetry parameter of a stationary state $\text{\cite{Kepesidis}}$, which provides a measure of the parity symmetry of the density operator, changes from 0 to a finite value at the transition point when $S=\infty$ (see Eq.(A.1) in the Supplemental Material $\text{\cite{supplemental}}$). This suggests the occurrence of the $\mathcal{PT}$ symmetry breaking of the stationary state $\text{\cite{Huber2}}$.

In addition, the eigenvalue structure and dynamics were obtained when $S=\infty$ $\text{\cite{Nakanishi}}$. In particular, in the $\mathcal{PT}$ phase, when $S=\infty$, the eigenvalues are expressed as 
\begin{align}
\label{2ptphase}
\lambda= in\sqrt{g^{2}-\Gamma^{2}}+r,
\end{align}
where $n\in\mathbb{N}$ and $r\in\mathbb{R}_{0}^{-}$, and information on degeneracy is omitted. Here, $\mathbb{R}_{0}^{-}$ is a set comprising real negative numbers including zero. Equation ($\ref{2ptphase}$) shows that commensurable pure imaginary eigenvalues exist. Therefore, some physical quantities such as magnetization oscillate periodically at late times. However, in the $\mathcal{PT}$ broken phase, all eigenvalues are real (see Eq.(A.3) in the Supplemental Material $\text{\cite{supplemental}}$); thus, the state decays toward a steady state without oscillation, which is the same as the stationary state. This behavior corresponds to one of the NHH $\mathcal{PT}$ phase transitions.

Moreover, it can be easily confirmed that $\textit{the $\mathcal{PT}$ phase is a boundary time crystal}$ because the eigenvalue structure satisfies the two conditions of nonstationary periodic dynamics only in the thermodynamic limit. (Detailed explanation of this model is provided in the Supplemental Material $\text{\cite{supplemental}}$.) In addition, if a model satisfies the definition ($\ref{HuberPT}$), it has been shown that there exists a stationary state that approaches the identity eigenmode $\rho\propto\1$ in the limit of zero dissipation rate $\text{\cite{Huber1}}$. \\

\textit{Boundary time crystals}. Among various BTC models, we first focused on the one-spin BTC model investigated in Refs.$\text{\cite{Iemini,Carmichael,Puri,Lawande,Hannukainen}}$. The Lindblad equation is expressed as
\begin{eqnarray}
\label{1spinBTC}
\frac{d}{dt}\rho=-2ig[S_{x},\rho]+\frac{\kappa}{S}\mathcal{D}[S_{-}]\rho,
\end{eqnarray}
where $g$ and $\kappa$ are the strengths of interaction and dissipation, respectively [Fig.$\ref{figponti}$ (b)].
In this model, the stationary state is solved exactly for finite $S$ $\text{\cite{Puri,Lawande}}$ as
\begin{eqnarray}
\label{ss}
\rho_{ss}=\frac{1}{D}\sum_{n,n^{\prime}=0}^{2S}\left(i\frac{\kappa}{g}\frac{S_{-}}{S}\right)^{n^{\prime}}\left(-i\frac{\kappa}{g}\frac{S_{+}}{S}\right)^{n},
\end{eqnarray}
where $D$ is the normalization constant.

In this model, it was found that various physical quantities, such as magnetization and purity, clearly change; namely, the dissipative phase transition occurs at $\kappa/g=1$ when $S\to\infty$ $\text{\cite{Puri,Lawande,Hannukainen}}$.

Moreover, the eigenvalue structure and dynamics were numerically investigated above and below the transition point $\text{\cite{Iemini}}$. For the BTC phase ($\kappa/g<1$), the real parts of many eigenvalues approach zero for an enormous $S$, and the imaginary parts are plotted at regular intervals for any $S$. This suggests that pure imaginary eigenvalues exist when $S\to\infty$ and that these eigenvalues are commensurable.
In other words, the imaginary part of the eigenvalues can be written as Im$[\lambda]=-icq$ when $S\to\infty$, where $q\in\mathbb{N}$ is the sector and $c$ is a real number dependent on $\kappa/g$. Here, the imaginary part is invariant within the same sector. However, for the BTC broken phase ($\kappa/g>1$), there are no eigenvalues with a non-zero imaginary part that approaches the imaginary axis as $S$ increases. \\

\textit{BTCs are $\mathcal{PT}$ symmetric phases}. Here, we first show that the one-spin BTC model ($\ref{1spinBTC}$) has the Liouvillian $\mathcal{PT}$ symmetry ($\ref{HuberPT}$). We choose the parity operator to reflect the basis of $S_z$, which acts on each spin operator as follows:
\begin{eqnarray}
\label{PSP}
PS_{z}P^{-1}=-S_{z},\ \ \ \ \ \ PS_{\pm}P^{-1}=S_{\mp}.
\end{eqnarray}

We can verify that $\mathbb{PT}(S_{x})= P\overline{S_{x}}P^{-1}=S_{x}$ and $\mathbb{PT^{\prime}}(S_{-})= P(S_{-})^{\dagger}P^{-1}=S_{-}$ hold, that is, the model has Liouvillian $\mathcal{PT}$ symmetry ($\ref{HuberPT}$).

Then, we analytically show that $\mathcal{PT}$ symmetry breaking of the stationary state $(\ref{ss})$ occurs at the BTC phase transition point. Using ($\ref{PSP}$), the conventional $\mathcal{PT}$ transformation of the stationary state can be written as: 
\begin{eqnarray}
\label{PTss}
PT\rho_{ss}PT=\frac{1}{D}\sum_{n,n^{\prime}=0}^{2S}\left(-i\frac{\kappa}{g}\frac{S_{+}}{S}\right)^{n}\left(i\frac{\kappa}{g}\frac{S_{-}}{S}\right)^{n^{\prime}}.
\end{eqnarray}
In the limit $S\to\infty$, we show that $\left(-i\frac{\kappa}{g}\frac{S_{+}}{S}\right)^{n}$ and $\left(i\frac{\kappa}{g}\frac{S_{-}}{S}\right)^{n^{\prime}}$ are commutative only when $\kappa/g<1$.

This implies that the {\it stationary state ($\ref{ss}$) of the one-spin BTC model is $\mathcal{PT}$ symmetric in the BTC phase but not in the BTC broken phase.} The details of the proof are provided in Sec. I.A of the Supplemental Material $\text{\cite{supplemental}}$. \\

Next, we consider the open one-spin model studied in Ref.$\text{\cite{Piccitto}}$ whose Lindblad equation is given by
\begin{align}
\label{ge1}
\frac{d}{dt}\rho=-i[H,\rho]+\frac{\kappa_{-}}{S}\mathcal{D}[S_{-}]\rho+\frac{\kappa_{+}}{S}\mathcal{D}[S_{+}]\rho,
\end{align}
where $H=S(g_{z}s_{z}^{p_{z}}+g_{x}s_{x}^{p_{x}})$,
and $p_{z}$, $p_{x}$ $\in\mathbb{N}$. and $s_{z},\ s_{x}$ are normalized spin operators $S_{z}/S,\ S_{x}/S$. 
This model has not been solved analytically; however, it has been numerically observed that BTCs appear when $p_{z}$ is even $\text{\cite{Piccitto}}$. By choosing the parity operator as a reflection of the basis of $S_{z}$ as before, the model can have Liouvillian $\mathcal{PT}$ symmetry ($\ref{HuberPT}$) if $p_{z}$ is even.

To investigate the $\mathcal{PT}$ symmetry breaking of a stationary state, we introduced the $\mathcal{PT}$-symmetry parameter $Q_{PT}$,
\begin{align}
Q_{PT}(\rho):=\frac{1}{Z}\sum_{i,j}|(\rho-PT\rho PT)_{ij}|,
\end{align}
where $Z:=\sum_{i,j}|(\rho)_{i,j}|+|(PT\rho PT)_{ij}|$ is a normalization constant, and thus $0\leq Q_{PT}\leq1$. Further, $(\rho)_{i,j}$ is the $(i,j) $ element of matrix $\rho$. If $Q_{PT}$ is zero, $\rho$ has $\mathcal{PT}$ symmetry. 

\begin{figure}[t]
   \vspace*{0.4cm}
     \hspace*{-5.3cm}
\includegraphics[bb=0mm 0mm 90mm 150mm,width=0.41\linewidth]{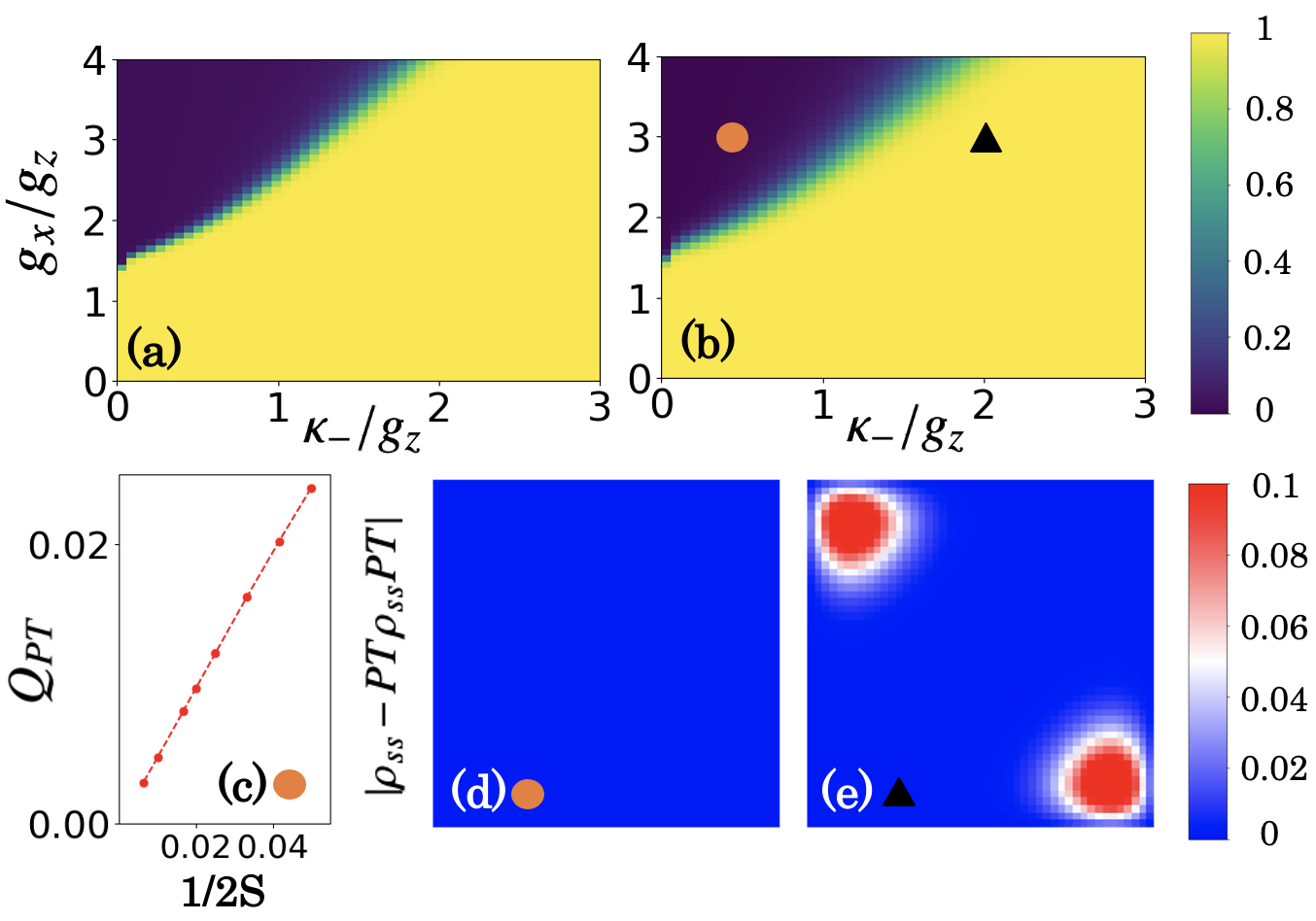}
  \caption{Numerical analysis of the model ($\ref{ge1}$) for $p_{z}=2,\ p_{x}=1$, $\kappa_{+}/g_{z}=0$, $S$=23. These are (a) purity, (b) $\mathcal{PT}$ symmetry parameter $Q_{PT}$, (c) $S$-dependence of $Q_{PT}$ for $\kappa_{-}/g_{z}=0.5$, $g_{x}/g_{z}=3$ (orange circle), (d) $|\rho_{ss}-PT\rho_{ss}PT|$ for $\kappa_{-}/g_{z}=0.5$, $g_{x}/g_{z}=3$ (orange circle), (e) $|\rho_{ss}-PT\rho_{ss}PT|$ for $\kappa_{-}/g_{z}=2$, $g_{x}/g_{z}=3$ (black triangle), where $|\rho |$ implies the matrix taking the absolute value for each element of the matrix $\rho$. Here, elements are computed on the basis of the $z$-magnetization. These results indicate that the stationary state has $\mathcal{PT}$ symmetry only in the BTC phase. Here, we have used QuTip $\text{\cite{Johansson1}}$ to obtain the stationary state numerically.}
    \label{fig1spingBTC}
\end{figure}

Figures $\rm\ref{fig1spingBTC}$ (a) and (b) show the purity and $\mathcal{PT}$-symmetry parameter $Q_{PT}$ of the stationary state for $p_{z}=2,\ p_{x}=1$. In the BTC phase (i.e., the phase with almost zero purity), $Q_{PT}$ is close to 0. Further, Fig.$\rm\ref{fig1spingBTC}$ (c) shows that $Q_{PT}$ decreases in the BTC phase with increase in $S$. Therefore, these results suggest that the $\mathcal{PT}$ symmetry of the stationary state is unbroken in the thermodynamic limit, whereas it is broken in the BTC broken phase. In addition, all elements of $|\rho_{ss}-PT\rho_{ss}PT|$ are close to 0 in the BTC phase [Fig.$\rm\ref{fig1spingBTC}$ (c)], whereas certain elements have finite values in the broken BTC phase (Fig.$\rm\ref{fig1spingBTC}$ (d)). Here, $|\rho |$ denotes the matrix that accepts the absolute value of each matrix element $\rho$. These results indicate that the stationary state exhibits $\mathcal{PT}$ symmetry only in the BTC phase. In Fig C.2 in the Supplemental Material $\text{\cite{supplemental}}$, we numerically investigated the time evolution and quantum trajectory of the normalized magnetization and normalized magnetization of the stationary state. \\

We also consider the one-spin model studied in Refs.$\text{\cite{Pedro,A,B}}$ whose Liouvillian is expressed as
\begin{align}
\label{Sz1}
\hat{\mathcal{L}}\rho=-ig[S_{x},\rho]+\frac{\kappa(1+p)}{S}\mathcal{D}[S_{x}^{+}]\rho+\frac{\kappa(1-p)}{S}\mathcal{D}[S_{x}^{-}]\rho,
\end{align}
where $-1\leq p\leq1$ and $S_{x}^{\pm}:=S_{y}\pm iS_{z}$. The model is BTC only when $p=0$. 
When $p=0$ the model is solvable for any $S$ $\text{\cite{Pedro}}$, the eigenvalues $\lambda_{l,q}$ are expressed as 
\begin{align}
\label{exact}
\lambda_{l,q}=igq - \frac{2\kappa}{S}[|q|+l(1+l+2|q|)],
\end{align}
where $q=\{-S,-S+1,..., S\}$ is the sector and $l=\{0,1,...,2S-|q|\}$, which satisfies the two conditions for the emergence of non-stationary oscillating dynamics only in the thermodynamic limit.

We can also discuss the relationship between the BTCs and $\mathcal{PT}$ symmetry for this model. 
Upon choosing the parity operator as the identity operator or reflection of the basis of $S_{z}$, this model ($\rm\ref{Sz1}$) satisfies Liouvillian $\mathcal{PT}$ symmetry only when $p=0$. This implies 
that {\it this model is BTC only when it has Liouvillian $\mathcal{PT}$ symmetry}. 
In addition, the stationary state $\rho_{ss}\propto\1$ has $\mathcal{PT}$ symmetry for $p=0$. 
In the following, we refer to this model with $p=0$ as the one-spin $\mathcal{PT}$ model [Fig.$\ref{figponti}$ (c)].\\

\textit{Understanding the mechanism of appearance of BTCs}. 
Let us focus the eigenvalues with the smallest real part for the one-spin $\mathcal{PT}$ model, namely $\lambda_{l=0,q}$ in Eq.($\rm\ref{exact}$). The corresponding eigenmodes $\rho_{0,q}$ are proportional to $(S_{x}^{+})^{|q|}$ for $q<0$ and $(S_{x}^{-})^{q}$ for $q>0$ $\text{\cite{B}}$. These exact specific eigenmodes facilitate an understanding of the BTC's mechanism.  For example, for $q=1$ we calculated $-ig[S_{x},S_{x}^{-}]=igS_{x}^{-}$ in the coherent part, and 
\begin{align}
&\frac{\kappa}{S}(\mathcal{D}[S_{x}^{+}]+\mathcal{D}[S_{x}^{-}])S_{x}^{-}=\frac{\kappa}{S}([S_{x}^{+},S_{x}^{-}]S_{x}^{-}+S_{x}^{-}[S_{x}^{-},S_{x}^{+}])\nonumber\\
&=\frac{2\kappa}{S}(S_{x}S_{x}^{-}-S_{x}^{-}S_{x})=\frac{2\kappa}{S}[S_{x},S_{x}^{-}]=-\frac{2\kappa}{S}S_{x}^{-},
\end{align}
in dissipative parts. [The calculation for any $q$ is provided in the Supplemental Material $\text{\cite{supplemental}}$; Eqs. (A.6), and (A.7).]
Moreover, it has a $1/S$-dependence in dissipative parts owing to the cancellation of several terms and the use of commutation relations. Thus, purely imaginary eigenvalues emerge when $p=0$ and $S\to\infty$, implying that they are BTC. 
However, when $p\neq0$, such cancellations of terms are generally unexpected. Indeed, the Liouvillian gap is not closed, even for $S\to\infty$ $\text{\cite{Pedro}}$ and no time crystals emerge. \\

Next, we investigated a class of the one-spin models using perturbation theory $\text{\cite{Fleming,Li}}$ under weak dissipations, whose Liouvillian is expressed as
\begin{align}
\label{Sx1}
\hat{\mathcal{L}}\rho&=-ig[S_{x},\rho]+\frac{\kappa}{S}\sum_{\mu}\mathcal{D}[L_{\mu}]\rho,\\
\label{Lmu}L_{\mu}&=\alpha_{\mu}S_{x}^{+}+\beta_{\mu}S_{x}^{-}+\gamma_{\mu}S_{x},
\end{align}
where $\alpha_{\mu},\ \beta_{\mu},\ \gamma_{\mu}\ \in\mathbb{C}$. This class includes the one-spin BTC model and the model ($\rm\ref{Sz1}$). We can show that BTCs appear in first-order perturbation under a weak dissipation rate $\kappa$ if and only if it holds that 
\begin{align}
\label{Lmup1}
\sum_{\mu}|\alpha_{\mu}|^{2}=\sum_{\mu}|\beta_{\mu}|^{2}.
\end{align}
Here, this condition can be regarded as exactly balanced total gain and loss on the $x$-basis. Note that BTCs basically appear under weak dissipations. Therefore, if it is not BTC in the first-order correction under weak dissipations, it may not be BTC for all dissipation regimes.

First, we apply the degenerated perturbation theory to the model ($\rm\ref{Sz1}$). For the prescription of the $n$-degenerate case, the first-order eigenvalue correction $\tilde{\lambda}_{n,i}^{(1)}$ can be obtained by solving the following equation: 
\begin{align}
\label{L}
L^{(n)}\psi_{n,i}=\tilde{\lambda}_{n,i}^{(1)}\psi_{n,i}\ \ \ \ \ \ \ (i=1,2,...,n),
\end{align}
where $L^{(n)}$ and $\psi_{n,i}$ are the $n$square matrix and $n$ coefficient vector, respectively. (see Eq.(D.10) in the Supplemental Material $\text{\cite{supplemental}}$).

Choosing the non-perturbative Liouvillian $\hat{\mathcal{L}}_0$ as a coherent part of the model ($\rm\ref{Sz1}$), namely $\hat{\mathcal{L}}_0\cdot =-ig[S_x,\cdot]$, the non-perturbative eigenvalues are ($2S+1-|q|$)-degenerated for each sector $q$.
Then, $L^{(2S+1-|q|)}$ is a tridiagonal matrix with real-number elements. In particular, for $p=0$, it becomes a symmetric matrix because of the balanced gain and loss (see Eq.(D.20) in the Supplemental Material $\text{\cite{supplemental}}$). 
In addition, all high-order corrections are zero because sector $q$ is invariant for the model ($\rm\ref{Sz1}$). Thus, perturbative analysis up to the first-order correction yields exact solutions.

\begin{figure}[t]
   \vspace*{0.8cm}
     \hspace*{-4.9cm}
\includegraphics[bb=0mm 0mm 90mm 150mm,width=0.45\linewidth]{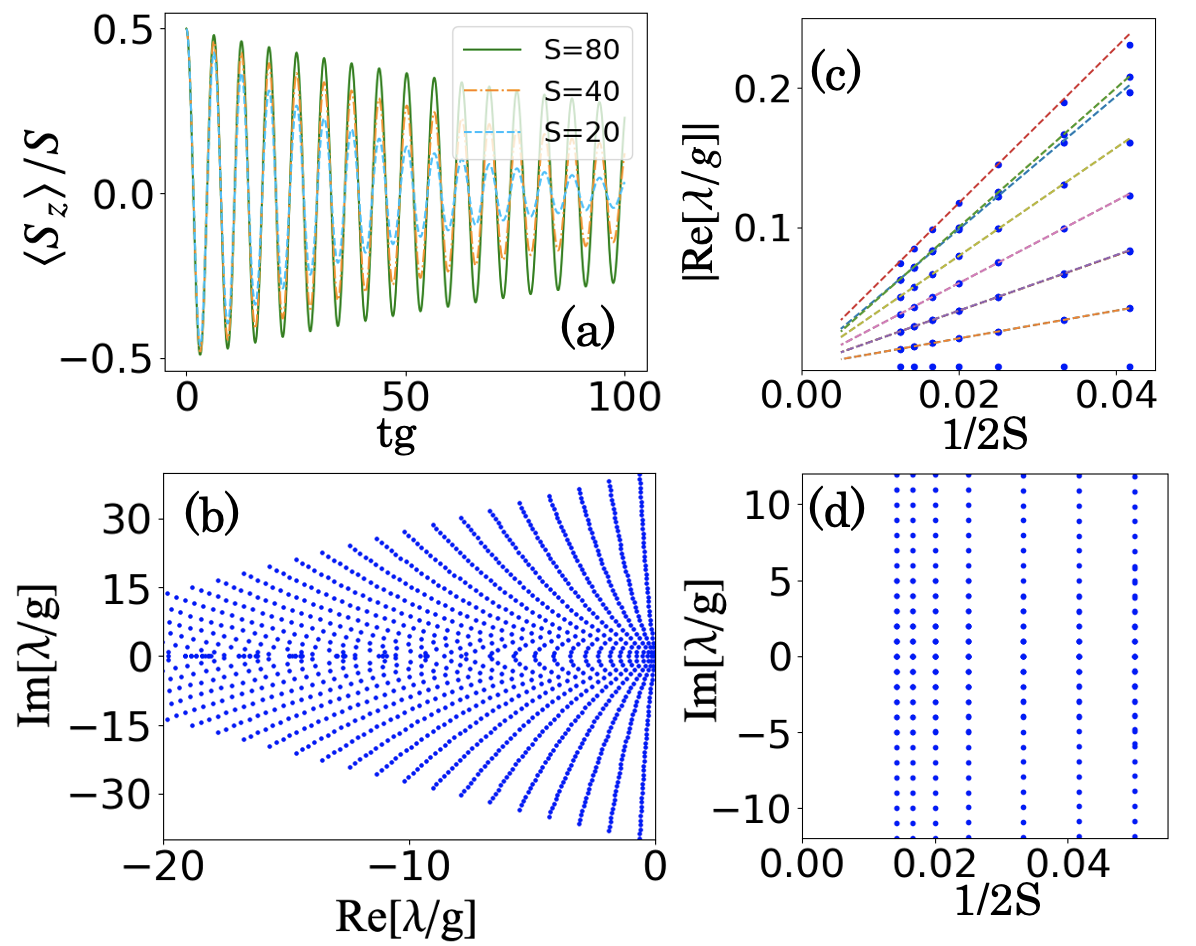}
    \caption{Numerical analysis of the time evolution of the normalized magnetization and Liouvillian spectrum in the model with $H=S_{x}$, $L=S_{z}$. (a) Time evolution of the normalized magnetization $\braket{S_{z}}/S$ for $S=20$ (dashed light blue), 40 (dashed-dotted orange), and 80 (solid green) and the initial state is set as $\rho(0)=\ket{S/2}_{z}\bra{S/2}_{z}$. (b) Liouvillian spectrum, (c) 15- minimum absolute values of real parts of the eigenvalues, (d) imaginary parts of the eigenvalues for $\kappa/g=1$ and $S=20$. Here, we have used QuTip $\text{\cite{Johansson1}}$ to obtain the Lindblad dynamics.}
    \label{figSz11}
\end{figure}

Considering these aspects, we also performed perturbation theory on a class of the one-spin models ($\rm\ref{Sx1}$). The non-perturbative Liouvillian $\hat{\mathcal{L}}_0$ was again chosen to be a coherent part, and consequently, the same tridiagonal real matrix $L^{(2S+1-|q|)}$ was obtained except for a constant multiplication and a sum of scalar multiplications. 
Therefore, its eigenvalues have the same properties, and the BTCs emerge for the symmetric case, but not for the non-symmetric case within the first-order perturbation scheme. In the Supplemental Material $\text{\cite{supplemental}}$, we provide details of our proof and numerically indicate that certain properties of the BTC phase transition can be caught up to the second-order corrections for the one-spin BTC model.

Choosing the parity operator as a reflection of the basis of $S_{z}$, the Liouvillian $\mathcal{PT}$ symmetry ($\rm\ref{HuberPT}$) guarantees the condition ($\rm\ref{Lmup1}$). Therefore, {\it model ($\rm\ref{Sx1}$) is BTC when it exhibits Liouvillian $\mathcal{PT}$ symmetry within the first-order perturbation scheme.} 
Note that conservation is not always true, that is, the condition ($\rm\ref{Lmup1}$) does not necessarily imply Liouvillian $\mathcal{PT}$ symmetry. This suggests that the definition of Liouvillian $\mathcal{PT}$ symmetry leaves room for further improvements.

Finally, we provide an example. When $L_{1}=-i(S_{x}^{+}-S_{x}^{-})/2=S_{z}$ in the model ($\ref{Sx1}$) that satisfies the condition ($\rm\ref{Lmup1}$), the normalized magnetization $\braket{S_{z}}/S$ oscillates and the relaxation time increases with increase in $S$ [Fig.$\rm\ref{figSz11}$ (a)]. Furthermore, Figs.$\rm\ref{figSz11}$ (c) and $\rm\ref{figSz11}$ (d) show that the real parts of the eigenvalues decrease to zero with an increase in $S$ and that the imaginary parts are invariant even as $S$ increases. These results imply that BTC emerges in the thermodynamic limit.

\textit{Summary and discussion}. In this study, we provided evidence that dissipative time crystals originating from $\mathcal{PT}$ symmetry exist in collective spin systems. In particular, we showed that BTCs are only such examples. In addition, we performed perturbation analysis for a class of one-spin models and showed that BTCs appear in the first-order correction because of the exactly balanced total gain and loss. 

Finally, we discuss the robustness of our results.
For a class of the one-spin models ($\rm\ref{Sx1}$), the BTCs were stable for perturbations of the dissipations satisfying Eq.($\rm\ref{Lmup1}$).
It was also stable in case of perturbations of Hamiltonian terms that do not break the Liouvillian $\mathcal{PT}$ symmetry ($\rm\ref{HuberPT}$), such as $S_z^{2n}$ or $S_x^n$ ($n\in\mathbb{N}$) $\text{\cite{Iemini,Piccitto}}$. However, the rigidity of the periodic time, which is a property of discrete time crystals, did not appear because the periodic time is generally the variant for the dissipation and interaction strength.

As a natural extension, investigation of the relationship between the Liouvillian $\mathcal{PT}$ symmetry and other time crystals, such as discrete time crystals, dissipative time crystals originating from dynamical symmetry, and boundary time crystals in bosonic systems are expected to yield interesting results.

\textit{Acknowledgment} - We thank Kohei Yamanaka and Yukiya Yanagihara for fruitful discussions. YN also acknowledges the financial support from JST SPRING, Grant Number JPMJSP2106, and Tokyo Tech Academy for Convergence of Materials and Informatics.
The work done by TS was supported by JSPS KAKENHI, Grants No. JP18H01141, No. JP18H03672, No. JP19L03665, No. JP21H04432, JP22H01143. Further, we would like to thank Editage (www.editage.com) for English language editing.

\onecolumngrid
\newpage
\Large{\centerline{Supplemental: \textit{Dissipative time crystals originating from parity-time symmetry}}}
\vskip\baselineskip
\large{\centerline{Yuma Nakanishi and Tomohiro Sasamoto}}

\renewcommand{\theequation}{A.\arabic{equation} }
\setcounter{equation}{0}


\section{Liouvillian $\mathcal{PT}$ symmetric models}
\renewcommand{\thefigure}{A.\arabic{figure} }
\setcounter{figure}{0}
\subsection{Two-spin model with gain and loss}
\normalsize{The Lindblad equation of the open two-spin-$S$ model with gain and loss $\text{\cite{Nakanishi,Huber1,Huber2}}$ is given by} 
\begin{align}
\label{2spind}
\frac{\partial}{\partial t}\rho=-ig[H,\rho]+\frac{\Gamma_{g}}{2S}\mathcal{D}[S_{+,A}]\rho+\frac{\Gamma_{l}}{2S}\mathcal{D}[S_{-,B}]\rho,
\end{align}
with $H=(S_{+,A}S_{-,B}+ \textrm{H.c.})/2S$.
This model satisfies the criterion of Huber et al. of Liouvillian $\mathcal{PT}$ symmetry ($\ref{HuberPT}$) when $\Gamma_{g}=\Gamma_{l}$ and the parity transformation is the exchange of two spins (i.e. $\mathbb{PT}(S_{+,A}S_{-,B}+ \textrm{H.c.})=(S_{+,A}S_{-,B}+ \textrm{H.c.})$, $\mathbb{PT}^{\prime}(S_{+,A})=S_{-,B}$, and $\mathbb{PT}^{\prime}(S_{-,B})=S_{+,A}$). Then we call it the two-spin $\mathcal{PT}$ model. Using the HP transformation $\text{\cite{Holstein}}$ and the third quantization $\text{\cite{Prosen3,Prosen4}}$, the physical quantities and eigenvalue structure can be obtained when $S=\infty$. In particular, for $\Gamma_{g}=\Gamma_{l}=\Gamma$, symmetry parameter $\Delta$ and purity, $\mu:=$ Tr$[\rho^{2}]$ in the stationary state are given by
\begin{align}
\Delta=\frac{|\braket{S_{+,A}S_{-,A}-S_{+,B}S_{-,B}}|}{\braket{S_{+,A}S_{-,A}+S_{+,B}S_{-,B}}}=&1-\left(\frac{g}{\Gamma}\right)^{2},\\
\mu=1-\left(\frac{g}{\Gamma}\right)^{2},&
\end{align}
for $\Gamma/g>1$, and $\Delta=\mu=0$ for $\Gamma/g<1$ when $S=\infty$.
Also, the Liouvillian eigenvalues $\lambda$ for each phase are given by
\begin{align}
\label{eigenPT}
\lambda=\begin{cases}
        {\ in\sqrt{g^{2}-\Gamma^{2}}+r, \ \ \ \ \ \ \ \ \ \ \ \ \ \ \ (\Gamma/g <1)},\\
        {\ -2(m_{+}\beta^{AFM}_{+}+m_{-}\beta^{AFM}_{-}), \ \ \ \ (\Gamma/g >1)},
    \end{cases}
\end{align}
where $r\in\mathbb{R}_{0}^{-}$, $n\in\mathbb{N}$, $m_{\pm}\in\mathbb{Z}^{+}$ and $\beta^{AFM}_{\pm}=(\Gamma\pm g)/2$ $\text{\cite{Nakanishi}}$. Note that the information on degeneracy is omitted in Eq.($\rm\ref{eigenPT}$). For $\Gamma/g<1$,  namely in the $\mathcal{PT}$ phase, we can easily confirm that (i) there exist pure imaginary eigenvalues $i\lambda_{j}$ and (ii) the quotient of each pure imaginary eigenvalue is a rational number only when $S=\infty$. $\textit{Therefore, we can find that the $\mathcal{PT}$ phase of the two-spin $\mathcal{PT}$ symmetric model is a boundary time crystal.}$

\begin{figure*}[b]
   \vspace*{-1.9cm}
     \hspace*{-10.6cm}
    \captionsetup{format=hang}
\includegraphics[bb=0mm 0mm 90mm 150mm,width=0.31\linewidth]{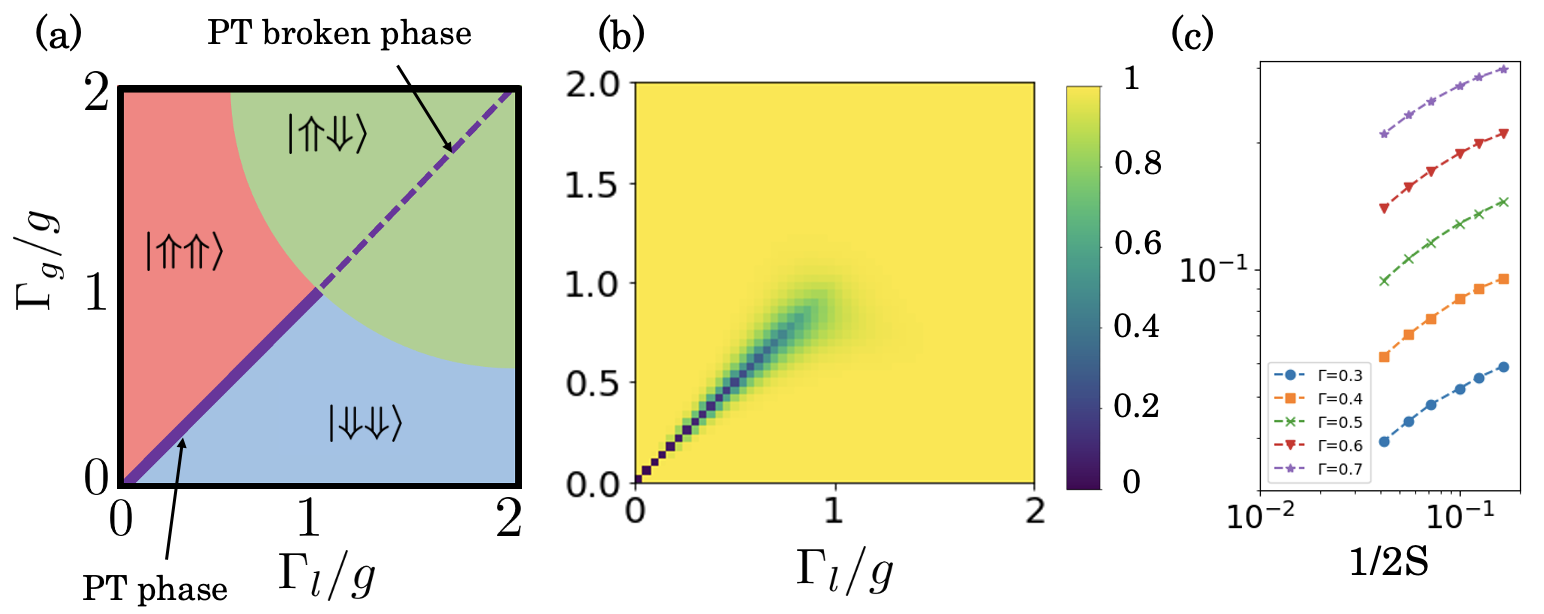}
    \caption{(a) Phase diagram of the open two-spin-$S$ model with gain and loss when $S\to\infty$ $\text{\cite{Nakanishi}}$. Red, blue, and green region indicates the ferromagnetic phase with $\braket{S_{z,A}/S}=\braket{S_{z,B}/S}=1$, the ferromagnetic phase with $\braket{S_{z,A}/S}=\braket{S_{z,B}/S}=-1$, the anti-ferromagnetic phase with $\braket{S_{z,A}/S}=1,\  \braket{S_{z,B}/S}=-1$, respectively. A real purple line indicates the $\mathcal{PT}$ phase, and a purple dashed line indicates the $\mathcal{PT}$ broken phase. (b) Numerical analysis for the $\mathcal{PT}$-symmetry parameter $Q_{PT}(\rho_{ss})$ for $S=7$. (c) the $S$-dependence of $Q_{PT}$ in the stationary state. These show that the $\mathcal{PT}$ symmetry breaking in the stationary state occurs in the thermodynamic limit.}
    \label{QPT2spin}
\end{figure*}

Let's mention here the properties of the symmetry parameter $\Delta$ and the $\mathcal{PT}$ symmetry parameter $Q_{PT}$. If $Q_{PT}(\rho)=0$, the density matrix $\rho$ is exactly $\mathcal{PT}$ symmetric. However, even if $\Delta(\rho)=0$, the density matrix $\rho$ is not always $\mathcal{PT}$ symmetric. Rather, it shows parity symmetry of the diagonal terms in the $S_{z}$ basis representation. If $\rho$ is Hermitian, the diagonal terms are real, and then $\Delta(\rho)=0$ is the necessary condition for the $\mathcal{PT}$ symmetry of the stationary state. 

Fig.$\rm\ref{QPT2spin}$ (a), (b) show the phase diagram when $S=\infty$ $\text{\cite{Nakanishi}}$ and the numerical calculation of the $\mathcal{PT}$ symmetry parameter $Q_{PT}$ in the stationary state for $S=7$. This shows that $Q_{PT}$ is less than 1 in the $\mathcal{PT}$ phase, while it is almost 1 in the $\mathcal{PT}$ broken phase. Fig.$\rm\ref{QPT2spin}$ (c) shows the $S$-dependence of $Q_{PT}$ in the stationary state for some dissipation strength. The symmetry parameter $Q_{PT}$ decreases as $S$ increases for each dissipation strength. These imply that the $\mathcal{PT}$ symmetry breaking in the stationary state occurs in the thermodynamic limit.


\subsection{One-spin $\mathcal{PT}$ model}
We consider the one-spin $\mathcal{PT}$ model whose Liouvillian is given by
\begin{align}
\label{1PT}
\hat{\mathcal{L}}\rho=-ig[S_{x},\rho]+\frac{\kappa}{S}\mathcal{D}[S_{x}^{+}]\rho+\frac{\kappa}{S}\mathcal{D}[S_{x}^{-}]\rho.
\end{align}
This model is equivalent to the model ($\rm\ref{Sz1}$) when $p=0$.
It has been exactly solved for any $S$ $\text{\cite{Pedro}}$, and the eigenvalues are given as Eq.($\ref{exact}$). In particular, for $l=0$, the eigenmodes $\rho_{0,q}$ are proportional to $(S_{x}^{+})^{|q|}$ for $q<0$ and $(S_{x}^{-})^{q}$ for $q>0$ $\text{\cite{B}}$. Indeed, substituting $(S_{x}^{-})^{q}$ for $\rho$ in the model ($\rm\ref{1PT}$), 
\begin{align}
\label{Szq1}
\hat{\mathcal{L}}(S_{x}^{-})^{q}&=-ig[S_{x},(S_{x}^{-})^{q}]+\frac{\kappa}{S}\mathcal{D}[S_{x}^{+}](S_{x}^{-})^{q}+\frac{\kappa}{S}\mathcal{D}[S_{x}^{-}](S_{x}^{-})^{q}\nonumber\\
&= iqg (S_{x}^{-})^{q}+\frac{\kappa}{S}(2S_{x}^{+}(S_{x}^{-})^{q}S_{x}^{-}-S_{x}^{-}S_{x}^{+}(S_{x}^{-})^{q}-(S_{x}^{-})^{q}S_{x}^{-}S_{x}^{+})
\nonumber\\
&\ \ \ \ \ \ \ \ \ \ \ \ \ \ \ \ \ \ \ \ \ +\frac{\kappa}{S}(2S_{x}^{-}(S_{x}^{-})^{q}S_{x}^{+}-S_{x}^{+}S_{x}^{-}(S_{x}^{-})^{q}-(S_{x}^{-})^{q}S_{x}^{+}S_{x}^{-})\nonumber\\
&= iqg (S_{x}^{-})^{q}+\frac{\kappa}{S}(S_{x}^{+}(S_{x}^{-})^{q}S_{x}^{-}-S_{x}^{-}S_{x}^{+}(S_{x}^{-})^{q}+S_{x}^{-}(S_{x}^{-})^{q}S_{x}^{+}-(S_{x}^{-})^{q}S_{x}^{+}S_{x}^{-})
\end{align}
Here, the several terms of dissipations are canceled out in Eq.($\rm\ref{Szq1}$). Further, it can be calculated as
\begin{align}
\label{Szq}
\hat{\mathcal{L}}(S_{x}^{-})^{q}&= iqg (S_{x}^{-})^{q}+\frac{\kappa}{S}([S_{x}^{+},S_{x}^{-}](S_{x}^{-})^{q}+(S_{x}^{-})^{q}[S_{x}^{-},S_{x}^{+}])
\nonumber\\
&= iqg (S_{x}^{-})^{q}+\frac{\kappa}{S}(2S_{x}(S_{x}^{-})^{q}-2(S_{x}^{-})^{q}S_{x})=iqg (S_{x}^{-})^{q}+\frac{2\kappa}{S}[S_{x},(S_{x}^{-})^{q}]\nonumber\\
&= iqg (S_{x}^{-})^{q}-\frac{2\kappa q}{S}(S_{x}^{-})^{q}= q(ig-\frac{2\kappa}{S})(S_{x}^{-})^{q},
\end{align}
where we use the commutation relations $[S_{x}^{\pm},S_{x}^{\mp}]=\pm2S_{x}$ and $[(S_{x}^{\pm})^{n},S_{x}]=\mp n(S_{x}^{\pm})^{n}$. 
Importantly, each term's order of spin operators decreases due to the commutation relations in Eq.($\rm\ref{Szq}$). These cause the real parts of the eigenvalues to have a dependence on $S^{-1}$, and then the pure imaginary eigenvalues emerge when $S\to\infty$.
The same argument also holds when substituting $(S_{x}^{+})^{|q|}$ for $\rho$ in the model ($\rm\ref{1PT}$). Also, such cancellations of terms are not expected in general. For example, the cancellations in Eqs.($\rm\ref{Szq1}$), ($\rm\ref{Szq}$) do not occur for the model ($\rm\ref{Sz1}$) when $p\neq0$, and then $(S_{x}^{-})^{q}$ does not become the eigenmode. Indeed Liouvillian gap is finite even for $S\to\infty$ $\text{\cite{Pedro,A}}$. It can be calculated by converting to a quadratic bosonic diagonalized Liouvillian using the Holstein-Primakoff approximation and the non-unitary Bogoliubov transform.

Furthermore, choosing the parity operator to be the identity operator or the reflection of the basis of $S_{z}$, this model ($\rm\ref{Sz1}$) satisfies the Liouvillian $\mathcal{PT}$ symmetry. Also, the stationary state $\rho_{ss}\propto\1$ has $\mathcal{PT}$ symmetry.

\section{Proof of the $\mathcal{PT}$ symmetry breaking in the stationary state for the one-spin BTC model}
\renewcommand{\theequation}{B.\arabic{equation} }
\setcounter{equation}{0}
We show that the stationary state $\rho_{ss}$ ($\ref{ss}$) has $\mathcal{PT}$ symmetry for $\kappa/g<1$, namely it satisfies $\rho_{ss}=PT\rho_{ss}PT$, while it is not $\mathcal{PT}$ symmetric for $\kappa/g>1$.
Firstly, we consider the case where $\kappa/g<1$.
Let us start from a well-known commutation relation of $S_{+}^{n}$ and $S_{-}$ $\text{\cite{Lawande}}$,
\begin{eqnarray}
\label{spinrelation}
S_{+}^{n}S_{-}=S_{-}S_{+}^{n}+n(n-1)S_{+}^{n-1}+2nS_{+}^{n-1}S_{z}.
\end{eqnarray}
By using Eq.($\rm{\ref{spinrelation}}$), we can write down the commutation relation of $S_{+}^{n}
$ and $S_{-}^{n^{\prime}}$ as
\begin{eqnarray}
\label{spinrelation2}
S_{+}^{n}S_{-}^{n^{\prime}}=S_{-}^{n^{\prime}}S_{+}^{n}+n(n-1)\sum_{k=1}^{n^{\prime}}S_{-}^{k-1}S_{+}^{n-1}S_{-}^{n^{\prime}-k}+2n\sum_{k=1}^{n^{\prime}}S_{-}^{k-1}S_{+}^{n-1}S_{z}S_{-}^{n^{\prime}-k}.
\end{eqnarray}
Therefore, the commutation relation of $
\left(-i\frac{\kappa}{g}\frac{S_{+}}{S}\right)^{n}$ and $\left(i\frac{\kappa}{g}\frac{S_{-}}{S}\right)^{n^{\prime}}$ in Eqs.($\ref{ss}$), ($\ref{PTss}$) can be written down as
\begin{eqnarray}
\label{spinrelation3}
\left[\left(-i\frac{\kappa}{g}\frac{S_{+}}{S}\right)^{n},\left(i\frac{\kappa}{g}\frac{S_{-}}{S}\right)^{n^{\prime}}\right]&=&+(-1)^{n}\frac{n(n-1)}{S^{2}}\left(i\frac{\kappa}{g}\right)^{n+n^{\prime}}\sum_{k=1}^{n^{\prime}}\left(\frac{S_{-}}{S}\right)^{k-1}\left(\frac{S_{+}}{S}\right)^{n-1}\left(\frac{S_{-}}{S}\right)^{n^{\prime}-k}\nonumber\\
&+&(-1)^{n}\frac{2n}{S}\left(i\frac{\kappa}{g}\right)^{n+n^{\prime}}\sum_{k=1}^{n^{\prime}}\left(\frac{S_{-}}{S}\right)^{k-1}\left(\frac{S_{+}}{S}\right)^{n-1}\left(\frac{S_{z}}{S}\right)\left(\frac{S_{-}}{S}\right)^{n^{\prime}-k}.
\end{eqnarray}
The first term on the right-hand side in Eq.($\rm\ref{spinrelation3}$) approaches 0 when $S$ $\to$ $\infty$ since it holds that
\begin{eqnarray}
\left|\left(\frac{\kappa}{g}\right)^{n+n^{\prime}}\frac{n(n-1)}{S^{2}}\sum_{k=1}^{n^{\prime}}\left(\frac{S_{-}}{S}\right)^{k-1}\left(\frac{S_{+}}{S}\right)^{n-1}\left(\frac{S_{-}}{S}\right)^{n^{\prime}-k}\right|\leq\left(\frac{\kappa}{g}\right)^{n+n^{\prime}}\frac{n(n-1)}{S^{2}}n^{\prime}e\to0\ \ \ (S\to\infty),
\end{eqnarray}
where we use that normalized spin operators are less than or equal to $\sqrt{1+1/S}$ and $(1+1/S)^{S}\leq e$ for $S>0$. Here, $e$ is Napier's constant.
Similarly, the second term on the right-hand side in Eq.($\rm\ref{spinrelation3}$) approaches 0 when $S$ $\to$ $\infty$. Therefore, we can show that $\lim_{S\to\infty}\left[\left(-i\frac{\kappa}{g}\frac{S_{+}}{S}\right)^{n}, \left(i\frac{\kappa}{g}\frac{S_{-}}{S}\right)^{n^{\prime}}\right]=0$ and then the stationary state ($\ref{ss}$) is $\mathcal{PT}$ symmetric when $S$ $\to$ $\infty$.

Next, we consider the case where $\kappa/g>1$. We transform the variable $\kappa/g$ as
\begin{eqnarray}
\frac{\kappa}{g}=\frac{1}{\sqrt{1-p^{2}}},
\end{eqnarray}
where $p$ can take a value from 0 to 1. Let us compare the elements $\bra{\ \floor{pS}\ }\rho_{ss}\ket{\ \floor{pS}\ }$ and $\bra{\ -\floor{pS}\ }\rho_{ss}\ket{\ -\floor{pS}\ }$, where the symbol $\floor{\ }$ means the floor function. We can calculate these elements as
\begin{eqnarray}
\label{floor}
\bra{\ \floor{pS}\ }\left(\frac{\kappa}{g}\right)^{2n}\left(\frac{S_{-}}{S}\frac{S_{+}}{S}\right)^{n}\ket{\ \floor{pS}\ }\simeq\prod_{l=0}^{n}\left(\frac{\kappa}{g}\right)^{2}\left[1-\left(\frac{\floor{pS}+l}{S}\right)^{2}\right]=\prod_{l=0}^{n}\frac{1-\left(\frac{\floor{pS}+l}{S}\right)^{2}}{1-p^{2}}
\end{eqnarray}
and 
\begin{eqnarray}
\label{floor1}
\bra{\ \floor{-pS}\ }\left(\frac{\kappa}{g}\right)^{2n}\left(\frac{S_{-}}{S}\frac{S_{+}}{S}\right)^{n}\ket{\ \floor{-pS}\ }\simeq\prod_{l=0}^{n}\left(\frac{\kappa}{g}\right)^{2}\left[1-\left(\frac{-\floor{pS}+l}{S}\right)^{2}\right]
=\prod_{l=0}^{n}\frac{1-\left(\frac{-\floor{pS}+l}{S}\right)^{2}}{1-p^{2}},
\end{eqnarray}
where we use the following relation, $\frac{S^{\pm}}{S}\ket{m}=\sqrt{\left(1\mp\frac{m}{S}\right)\left(1\pm\frac{m}{S}+\frac{1}{S}\right)}\ket{m\pm1}\simeq\sqrt{1-\left(\frac{m}{S}\right)^{2}}\ket{m\pm1}$ for a large $S$.
We find that Eq.($\rm\ref{floor}$) is less than or equal to Eq.($\rm\ref{floor1}$) since $\left(\floor{pS}+l\right)^{2}\geq\left(-\floor{pS}+l\right)^{2}$. In particular, for $n=2\floor{pS}$, Eq.($\rm\ref{floor}$) is less than Eq.($\rm\ref{floor1}$). As a result, it holds that $\bra{\ \floor{pS}\ }\rho_{ss}\ket{\ \floor{pS}\ }<\bra{\ -\floor{pS}\ }\rho_{ss}\ket{\ -\floor{pS}\ }$ for $0<p<1$, and then it can be seen that the $\mathcal{PT}$ symmetry of the steady state $\rho_{ss}$ ($\ref{ss}$) is broken for $\kappa/g>1$ when $S\to\infty$.


\section{Numerical calculation for the BTC models}
\renewcommand{\theequation}{C.\arabic{equation} }
\setcounter{equation}{0}
\renewcommand{\thefigure}{C.\arabic{figure} }
\setcounter{figure}{0}

\subsection{Numerical calculation for the one-spin BTC model}

\begin{figure*}[htbp]
   \vspace*{0.6cm}
     \hspace*{-6.1cm}
    \captionsetup{format=hang}
\includegraphics[bb=0mm 0mm 90mm 150mm,width=0.27\linewidth]{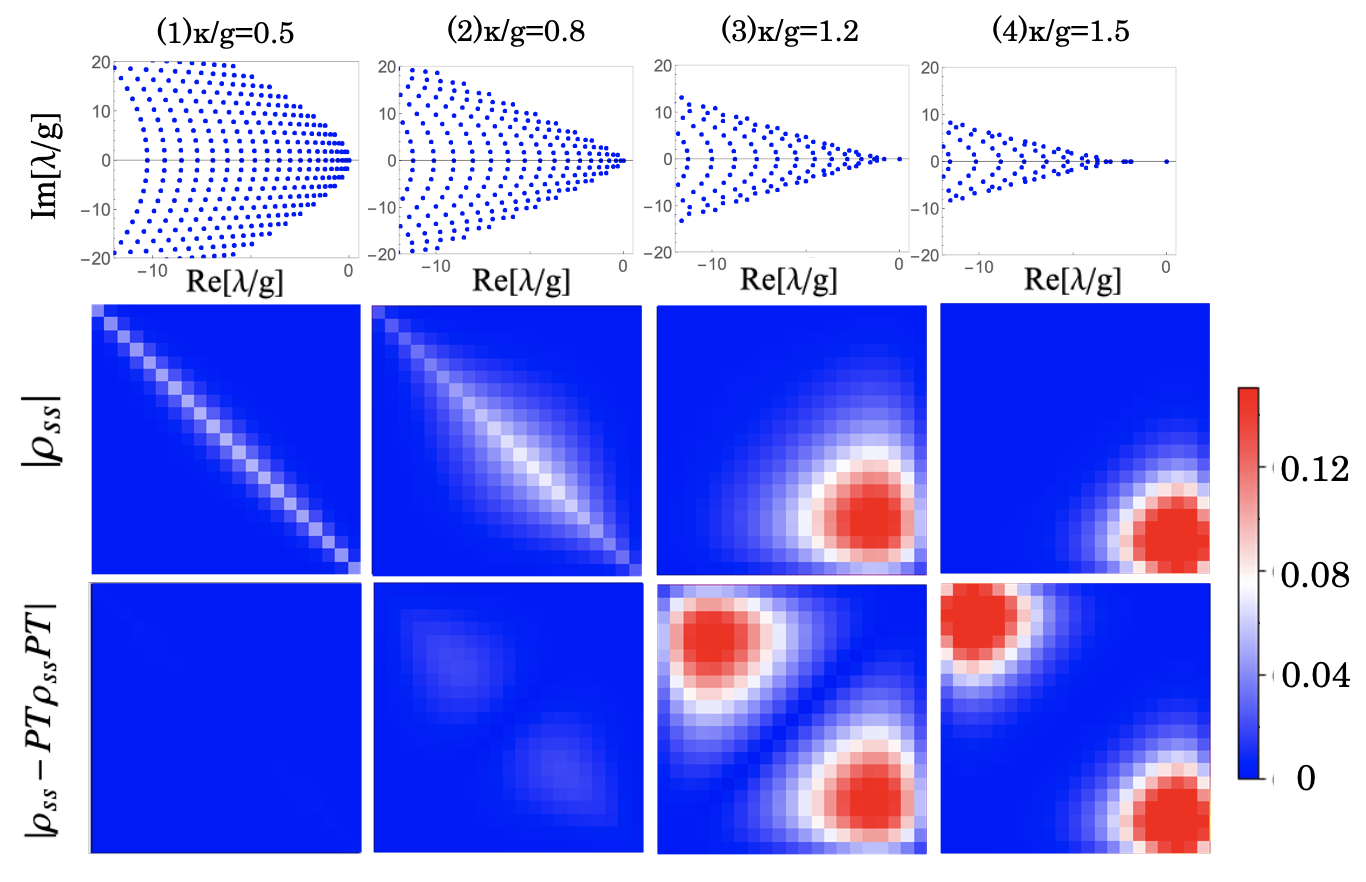}
    \caption{Numerical analysis of the one-spin BTC model for $S=10$ at $\kappa/g=0.5,\ 0.8,\ 1.2,\ 1.5$. Top: Eigenvalue structures\ \  Medium: $|\rho_{ss}|$\ \  Bottom: $|\rho_{ss}-PT\rho_{ss}PT|$. Here, $|\rho |$ means the matrix taking the absolute value for each element of the matrix $\rho$, and the elements are computed
on the basis of the $z$-magnetization. These results show that the stationary state $\rho_{ss}$ is $\mathcal{PT}$ symmetric in the BTC phase, while it is not in the BTC broken phase.}
    \label{fig1spinBTC}
\end{figure*}

We numerically investigate the $\mathcal{PT}$ symmetry breaking of the stationary state. 
Fig.$\rm\ref{fig1spinBTC}$ shows eigenvalue structures (top) and $|\rho_{ss}|$ (medium) and $|\rho_{ss}-PT\rho_{ss}PT|$ (bottom) for $S=10$. Here, $|\rho |$ means the matrix takes the absolute value for each matrix element $\rho$. The top figures show that there exist near pure imaginary numbers in the BTC phase, while eigenvalues with the slow decay eigenmodes are real in the BTC broken phase.
Medium figures imply that the stationary state $\rho_{ss}$ is likely to be $\mathcal{PT}$ symmetric in the BTC phase, while it is broken in the BTC broken phase. Also, the bottom figures indicate that all the elements of $|\rho_{ss}-PT\rho_{ss}PT|$ are close to 0 in the BTC phase, while some elements are finite values in the BTC broken phase. These results indicate that the stationary state $\rho_{ss}$ is $\mathcal{PT}$ symmetric in the BTC phase, while it is not in the BTC broken phase.

\subsection{Numerical calculation of the model ($\ref{ge1}$)}
\begin{figure*}[htbp]
   \vspace*{2.9cm}
     \hspace*{-10.2cm}
    \captionsetup{format=hang}
\includegraphics[bb=0mm 0mm 90mm 150mm,width=0.28\linewidth]{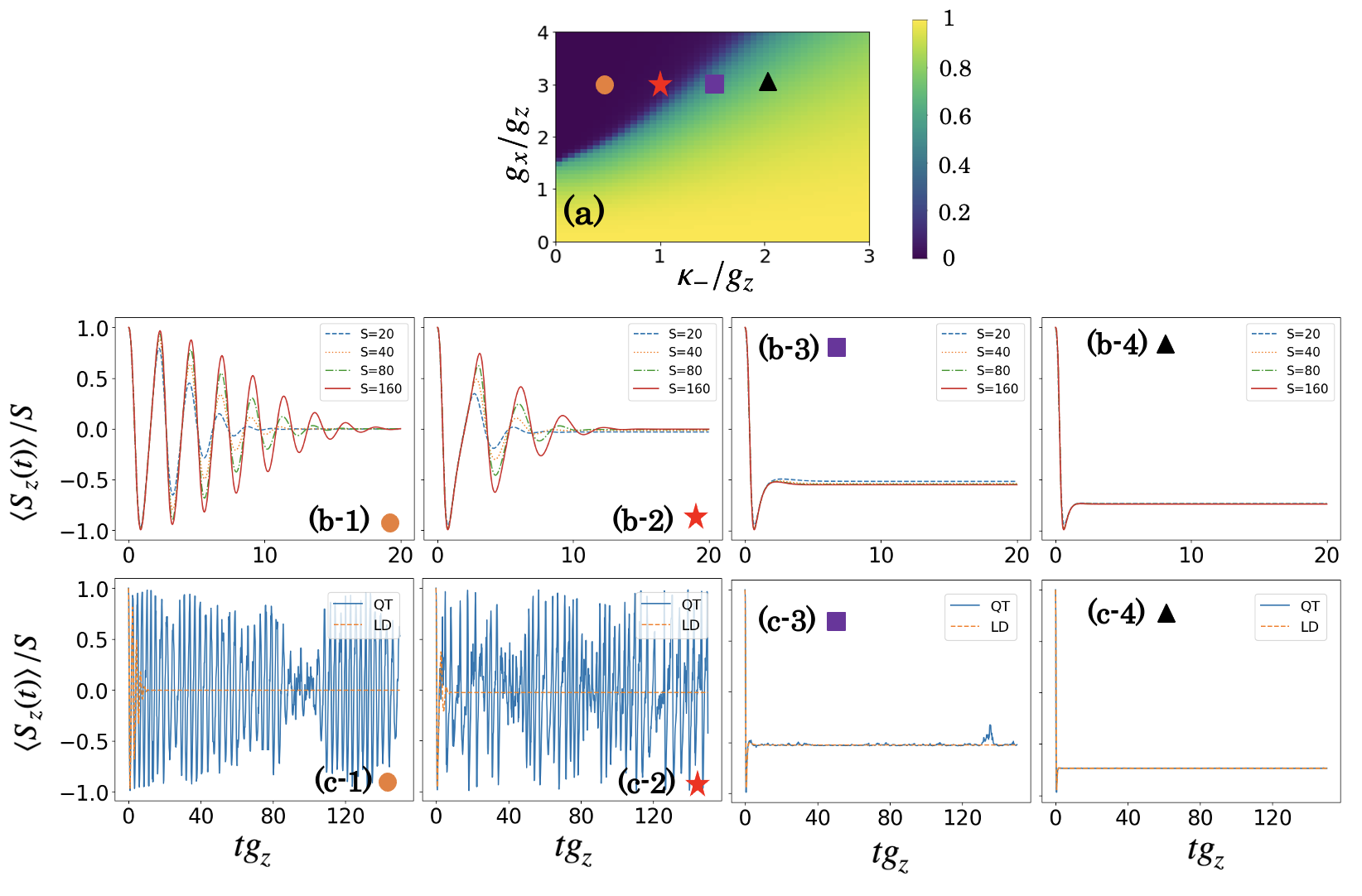}
    \caption{Numerical analysis of the model ($\ref{ge1}$) for $p_{z}=2,\ p_{x}=1$, $\kappa_{+}/g_{z}=0$. (a) The normalized magnetization of the stationary state $\braket{S_z}/S$ for $S$=23. (b) The time evolution with Lindblad dynamics (LD) for $S$=20,40,80,160, and (c) the quantum trajectory (QT) for $S$=23 of the normalized magnetization for $g_{x}/g_{z}=3$, and (b-1), (c-1) $\kappa_{-}/g_{z}=0.5$ (orange circle), (b-2), (c-2) $\kappa_{-}/g_{z}=1$ (red star), (b-3), (c-3) $\kappa_{-}/g_{z}=1.5$ (purple square), (b-4), (c-4) $\kappa_{-}/g_{z}=2$ (black triangle). These results imply that the BTC phase can be determined by the dynamics, the quantum fluctuations, or the properties of the stationary state, such as purity, magnetization, and $\mathcal{PT}$ symmetry parameter $Q_{PT}$. Here, we have used QuTip $\text{\cite{Johansson1}}$ to numerically obtain the stationary state and the quantum trajectory.}
    \label{1spinggBTC}
\end{figure*}

BTC and Liouvillian $\mathcal{PT}$ phases can be determined not only by dynamical properties such as dynamics $\text{\cite{Iemini,Piccitto,dos,Carmichael,Huber2,Nakanishi}}$ and quantum trajectory $\text{\cite{Nakanishi,Link}}$ but also by static properties such as magnetization $\text{\cite{Iemini,Huber2,Huber1,Nakanishi,Hannukainen,Puri,Lawande}}$, purity $\text{\cite{Piccitto,Huber1,Hannukainen}}$, $\mathcal{PT}$ symmetry parameter $Q_{PT}$ of the stationary state.
Here we numerically investigate the normalized magnetization of the stationary state, the time evolution, and the quantum trajectory of the normalized magnetization for the model ($\ref{ge1}$) with $p_{z}=2,\ p_{x}=1$. 
Fig.$\rm\ref{1spinggBTC}$ (a) shows the magnetization of the stationary state. Magnetization can usually be regarded as the order parameter even in dissipative systems. Comparing Fig.$\rm\ref{fig1spingBTC}$ (a), (b), we can see that the magnetization is also zero in the region where the purity and $Q_{PT}$ are zero.
Next, we investigate the time evolution with fixed $g_x/g_{z}$ in Fig.$\rm\ref{1spinggBTC}$ (b). These results show that in the BTC phase, the magnetization periodically oscillates, and the relaxation time increases with increasing $S$. On the other hand, in the BTC broken phase, the magnetization decays without oscillation, and the behavior of dynamics little changes with increasing $S$.
Next, we examine the quantum trajectory at the same point in Fig.$\rm\ref{1spinggBTC}$ (c). In the BTC (Liouvillian $\mathcal{PT}$) phase, quantum fluctuations are known to be large since the contribution from quantum jumps is dominant due to exactly balanced dissipation $\text{\cite{Nakanishi,Link}}$.
Indeed, these results show that the fluctuations are large in the BTC phase and small in the BTC phase.

\section{Perturbation theory for the one-spin models}
\renewcommand{\theequation}{D.\arabic{equation} }
\setcounter{equation}{0}
\renewcommand{\thefigure}{D.\arabic{figure} }
\setcounter{figure}{0}

\subsection{Degenerate perturbation theory of Liouvillians}
We consider the perturbation theory of Liouvillians $\text{\cite{Fleming,Li}}$. Firstly, a Liouvillian $\hat{\mathcal{L}}$ is divided into the non-perturbative part $\hat{\mathcal{L}}_{0}$ and the perturbative part $\hat{\mathcal{L}}_{1}$,
\begin{align}
\hat{\mathcal{L}}(\alpha)=\hat{\mathcal{L}}_{0}+\alpha \hat{\mathcal{L}}_{1}.
\end{align}
Suppose that the eigenmodes of the non-perturbative part $\hat{\mathcal{L}}_{0}$ are $u_n^{(0)}$ with eigenvalue $\lambda_{n}^{(0)}$, namely it holds that $\hat{\mathcal{L}}_{0}u_n^{(0)}=\lambda_{n}^{(0)}u_n^{(0)}$, and the eigenmodes of the Liouvillian $\hat{\mathcal{L}}(\alpha)$ are $u_n(\alpha)$ with eigenvalue $\lambda_{n}(\alpha)$, namely it holds that $\hat{\mathcal{L}}(\alpha)u_{n}(\alpha)=\lambda_{n}(\alpha)u_{n}(\alpha)$. The Hilbert-Schmidt inner product is introduced as $\braket{\rho,\sigma}:=\text{Tr}[\rho^{\dagger}\sigma]$, and the Hermitian adjoint of the Liouvillian $\hat{\mathcal{L}}^{\dagger}$ is also defined as 
\begin{align}
\braket{\hat{\mathcal{L}}\rho,\sigma}=\braket{\rho,\hat{\mathcal{L}}^{\dagger}\sigma}.
\end{align}
Then it holds that
\begin{align}
\hat{\mathcal{L}}^{\dagger}\omega_{n}=(\lambda_{n})^{*}\omega_{n},\ \ \ \ \ \ \ \ \ \ \ \ \braket{\omega_{m},u_{n}}=\delta_{mn}.
\end{align}

Next, the eigenvalues and eigenmodes are expanded as 
\begin{align}
u_{n}(\alpha)=\sum_{k=0}^{\infty}\alpha^{k}u_{n}^{(k)},\ \ \ \ \ \ \ \ 
\lambda_{n}(\alpha)=\sum_{k=0}^{\infty}\alpha^{k}\lambda_{n}^{(k)},
\end{align}
where we assume that eigenvalues do not degenerate.
Also, using the conventional perturbation prescription, the following equation can be obtained, 
\begin{align}
\lambda_{n}^{(1)}=\braket{\omega_{n}^{(0)},\hat{\mathcal{L}}_{1}u_{n}^{(0)}},\\
u_{n}^{(1)}=\sum_{k\neq n}\frac{\braket{\omega_{k}^{(0)},\hat{\mathcal{L}}_{1}u_{n}^{(0)}}}{\lambda_{n}^{(0)}-\lambda_{k}^{(0)}}u_{k}^{(0)},\\
\lambda_{n}^{(2)}=\sum_{k\neq n}\frac{\braket{\omega_{n}^{(0)},\hat{\mathcal{L}}_{1}u_{k}^{(0)}}\braket{\omega_{k}^{(0)},\hat{\mathcal{L}}_{1}u_{n}^{(0)}}}{\lambda_{n}^{(0)}-\lambda_{k}^{(0)}}.
\end{align}

Next, we consider the degenerate case. In this case, new non-perturbative eigenmodes are constructed as 
\begin{align}
\tilde{u}_{n,i}^{(0)}=\sum_{j}c_{ji}u_{n,j}^{(0)},
\end{align}
with
\begin{align}
\braket{\tilde{\omega}_{n,j}^{(0)},\hat{\mathcal{L}}_{1}\tilde{u}_{n,i}^{(0)}}=0 \ \ \ \ \ \textrm{for}\ \ \  i\neq j,
\end{align}
where $u_{n,j}^{(0)}$ is a non-perturbative eigenmode with eigenvalue $\lambda_{n}^{(0)}$. Then, the coefficient $c_{ji}$ and the first-order eigenvalue correction $\tilde{\lambda}_{n}^{(1)}$ can be given by solving the following secular equation,
\begin{eqnarray}
\label{dmat}
\left(
    \begin{array}{cccc}
      \braket{1,\hat{\mathcal{L}}_{1}1} & \braket{1,\hat{\mathcal{L}}_{1}2}   & \cdots& \cdots \\
     \braket{2,\hat{\mathcal{L}}_{1}1}   & \braket{2,\hat{\mathcal{L}}_{1}2}  &\cdots&\cdots \\
      \vdots & \vdots&\cdots  & \cdots \\
     \vdots &\vdots & \cdots& \cdots
    \end{array}
 \right)\left(
    \begin{array}{c}
      c_{1i} \\
     c_{2i} \\
      \vdots\\
     \vdots 
    \end{array}
 \right)=\tilde{\lambda}_{n,i}^{(1)}\left(
    \begin{array}{c}
      c_{1i} \\
     c_{2i} \\
      \vdots\\
     \vdots 
    \end{array}
 \right),
\end{eqnarray}
where $\braket{i,\hat{\mathcal{L}}_{1}j}:=\braket{\omega_{n,i}^{(0)},\hat{\mathcal{L}}_{1}u_{n,j}^{(0)}}$. Also, the first-order eigenmodes correction and the second-order eigenvalues correction are given by 
\begin{align}
\tilde{u}_{n,i}^{(1)}=\sum_{k\neq n}\sum_{j}\frac{\braket{\tilde{\omega}_{k,j}^{(0)},\hat{\mathcal{L}}_{1}\tilde{u}_{n,i}^{(0)}}}{\lambda_{n}^{(0)}-\lambda_{k}^{(0)}}\tilde{u}_{k,j}^{(0)},\\
\label{2dp}
\tilde{\lambda}_{n}^{(2)}=\sum_{k\neq n}\sum_{j}\frac{\braket{\tilde{\omega}_{n,i}^{(0)},\hat{\mathcal{L}}_{1}\tilde{u}_{k,j}^{(0)}}\braket{\tilde{\omega}_{k,j}^{(0)},\hat{\mathcal{L}}_{1}\tilde{u}_{n,i}^{(0)}}}{\lambda_{n}^{(0)}-\lambda_{k}^{(0)}}.
\end{align}

\subsection{Perturbation theory for the model ($\ref{Sz1}$)}
We analyze the one-spin model ($\ref{Sz1}$) using the (degenerate) perturbation theory. Now, we choose that the non-perturbative part is the coherent part in Eq.($\rm\ref{Sz1}$), namely $\hat{\mathcal{L}}_{0}[\cdot]=-i[H,\cdot]$, and the perturbative parts are dissipation parts in Eq.($\rm\ref{Sz1}$), namely $\hat{\mathcal{L}}_{1}=\hat{\mathcal{L}}_{+}(1+p)/S+\hat{\mathcal{L}}_{-}(1-p)/S,$ where $\hat{\mathcal{L}}_{+}=\mathcal{D}[S_{x}^{+}]$ and $\hat{\mathcal{L}}_{-}=\mathcal{D}[S_{x}^{-}]$. Here, $\kappa$ is the perturbation parameter. In this case, $u^{(0)}$ and $\omega^{(0)}$ and eigenvalues are written as 
\begin{align}
\label{uw}
u^{(0)}=\omega^{(0)}=\rho_{n,q}=\ket{n}_{x}\bra{n-q}_{x},\ \ \ \ \ \ \ \ \lambda_{n,q}^{(0)}=-iq,
\end{align}
where the subscript $x$ of the bra and ket means the eigenbasis of the operator $S_{x}$. It can be easily found that the eigenvalues $-iq$ are $(2S+1-|q|)$-order degenerated. Also, the Hilbert-Schmidt inner products $\braket{\rho_{m,q^{\prime}},\hat{\mathcal{L}}_{+}\rho_{n,q}}$ and $\braket{\rho_{m,q^{\prime}},\hat{\mathcal{L}}_{-}\rho_{n,q}}$ can be calculated as
\begin{align}
\label{+}
\braket{\rho_{m,q^{\prime}},\hat{\mathcal{L}}_{+}\rho_{n,q}}&=2\sqrt{(S-n)(S+n+1)(S-n+q)(S+n-q+1)}\delta_{m,n+1}\delta_{q^{\prime},q}\nonumber\\
&-\{(S-n)(S+n+1)+(S-n+q)(S+n-q+1)\}\delta_{m,n}\delta_{q^{\prime},q},\\
\label{-}
\braket{\rho_{m,q^{\prime}},\hat{\mathcal{L}}_{-}\rho_{n,q}}&=2\sqrt{(S+n)(S-n+1)(S+n-q)(S-n+q+1)}\delta_{m,n-1}\delta_{q^{\prime},q}\nonumber\\
&-\{(S+n)(S-n+1)+(S+n-q)(S-n+q+1)\}\delta_{m,n}\delta_{q^{\prime},q}.
\end{align}
Therefore, the matrices in Eq.($\rm\ref{dmat}$) for the sector $q$ can be written as the $(2S+1-|q|)$ real tridiagonal matrices on the basis of the $x$-magnetization with the elements
\begin{align}
\label{16}
\braket{\rho_{n,q},\hat{\mathcal{L}}_{1}\rho_{n,q}}&=-\frac{1+p}{S}\{(S-n)(S+n+1)+(S-n+q)(S+n-q+1)\}\nonumber\\
&-\frac{1-p}{S}\{(S+n)(S-n+1)+(S+n-q)(S-n+q+1)\},\\
\label{17}
\braket{\rho_{n+1,q},\hat{\mathcal{L}}_{1}\rho_{n,q}}&=\frac{2(1+p)}{S}\sqrt{(S-n)(S+n+1)(S-n+q)(S+n-q+1)},\\
\label{18}
\braket{\rho_{n-1,q},\hat{\mathcal{L}}_{1}\rho_{n,q}}&=\frac{2(1-p)}{S}\sqrt{(S+n)(S-n+1)(S+n-q)(S-n+q+1)},
\end{align}
with $n=\{S, S-1, \cdots, -S+|q|\}$.
For $p=0$, all the matrices are symmetric, since
\begin{align}
\braket{\rho_{n,q},\hat{\mathcal{L}}_{1}\rho_{n-1,q}}=\braket{\rho_{n-1,q},\hat{\mathcal{L}}_{1}\rho_{n,q}},
\end{align}
while for $p\neq0$, they are not symmetric.

For example, $p=q=0$, using Eqs.($\rm\ref{16}$)-($\rm\ref{18}$), the matrix in Eq.($\rm\ref{dmat}$) can be written as the $(2S+1)$ symmetric tridiagonal matrix in the basis of the $x$-magnetization
\begin{eqnarray}
\label{dmat2}
4\left(
    \begin{array}{ccccc}
      -1& 1  &0 & \cdots& \cdots \\
    1   & -3+\frac{1}{S}& 2-\frac{1}{S} &\cdots&\cdots \\
     0  & 2-\frac{1}{S}&  -5+\frac{4}{S} &\cdots&\cdots \\
      \vdots & \vdots&\vdots  & \ddots& \ddots \\
     \vdots &\vdots & \vdots&\ddots& \ddots
    \end{array}
 \right),
\end{eqnarray}
where we use the following relations,
\begin{align}
\braket{\rho_{n,0},\hat{\mathcal{L}}_{1}\rho_{n,0}}=-\frac{4}{S}(S^{2}+S-n^{2}),\\
\braket{\rho_{n+1,0},\hat{\mathcal{L}}_{1}\rho_{n,0}}=\frac{2}{S}(S-n)(S+n+1),\\
\braket{\rho_{n-1,0},\hat{\mathcal{L}}_{1}\rho_{n,0}}=\frac{2}{S}(S+n)(S-n+1),
\end{align}
with $n=\{S, S-1, \cdots, -S\}$. Diagonalizing the matrix ($\rm\ref{dmat2}$), it can be found that many first-order eigenvalue corrections are closer to 0 as $S$ increases, and all the eigenvalues are real. This means that many eigenvalues are zero when $S\to\infty$.


In this model, all the high-order perturbation terms are 0 since $\braket{\tilde{\rho}_{m,q^{\prime}},\hat{\mathcal{L}}_{1}\tilde{\rho}_{n,q}}=0$, for $(m,q^{\prime})\neq(n,q)$ since the sector $q$ is invariant when $\hat{\mathcal{L}}_{1}$ acts on $\tilde{\rho}_{n,q}$ as shown in Eq.($\rm\ref{+}$), ($\rm\ref{-}$). So the perturbative analysis up to first-order correction gives the exact result for the one-spin model ($\rm\ref{Sz1}$).

\subsection{Perturbation theory for a class of the one-spin model}
Next, we apply the degenerate perturbation theory to a class of the one-spin model ($\rm\ref{Sx1}$). Now, we choose that the non-perturbative part is a coherent part in Eq.($\rm\ref{Sx1}$), namely $\hat{\mathcal{L}}_{0}[\cdot]=-i[H,\cdot]$, and the perturbative part is the dissipation part in Eq.($\rm\ref{Sx1}$), namely $\hat{\mathcal{L}}_{1}=\sum_{\mu}\hat{\mathcal{L}}_{\mu}/S=\sum_{\mu}\mathcal{D}[L_{\mu}]/S$. Here, $\kappa$ is the perturbation parameter. In this case, $u^{(0)}$ and $\omega^{(0)}$ and eigenvalues are also written as Eq.($\rm\ref{uw}$). Now, we calculate the Hilbert-Schmidt inner product $\braket{\rho_{m,q},\hat{\mathcal{L}}_{\mu}\rho_{n,q}}$,
\begin{align}
\label{lmu}
\braket{\rho_{m,q},\hat{\mathcal{L}}_{\mu}\rho_{n,q}}&=2|\alpha_{\mu}|^{2}\sqrt{(S-n)(S+n+1)(S-n+q)(S+n-q+1)}\delta_{m,n+1}\nonumber\\
&+2|\beta_{\mu}|^{2}\sqrt{(S+n)(S-n+1)(S+n-q)(S-n+q+1)}\delta_{m,n-1}\nonumber\\
&-|\alpha_{\mu}|^{2}\{(S-n)(S+n+1)+(S-n+q)(S+n-q+1)\}\delta_{m,n}\nonumber\\
&-|\beta_{\mu}|^{2}\{(S+n)(S-n+1)+(S+n-q)(S-n+q+1)\}\delta_{m,n}-|\gamma_{\mu}|^{2}q^{2}\delta_{m,n}.
\end{align}
Note that there exists non-zero Hilbert-Schmidt inner product $\braket{\rho_{m,q^{\prime}},\hat{\mathcal{L}}_{\mu}\rho_{n,q}}$ for $q^{\prime}\neq q$ in general, so the high-order perturbation terms are not 0 in this class. However, these terms do not contribute to the first-order correction.

Similarly to the model ($\rm\ref{Sz1}$), the matrices in Eq.($\rm\ref{dmat}$) for the sector $q$ can be written as the $(2S+1-|q|)$ real tridiagonal matrices on the basis of the $x$-magnetization with the elements
\begin{align}
\label{25}
\braket{\rho_{n,q},\hat{\mathcal{L}}_{1}\rho_{n,q}}&=-\sum_{\mu}\frac{|\alpha_{\mu}|^{2}}{S}\{(S-n)(S+n+1)+(S-n+q)(S+n-q+1)\}\nonumber\\
&-\frac{|\beta_{\mu}|^{2}}{S}\{(S+n)(S-n+1)+(S+n-q)(S-n+q+1)\}-\frac{|\gamma_{\mu}|^{2}q^{2}}{S},\\
\label{26}
\braket{\rho_{n+1,q},\hat{\mathcal{L}}_{1}\rho_{n,q}}&=\sum_{\mu}\frac{2|\alpha_{\mu}|^{2}}{S}\sqrt{(S-n)(S+n+1)(S-n+q)(S+n-q+1)},\\
\label{27}
\braket{\rho_{n-1,q},\hat{\mathcal{L}}_{1}\rho_{n,q}}&=\sum_{\mu}\frac{2|\beta_{\mu}|^{2}}{S}\sqrt{(S+n)(S-n+1)(S+n-q)(S-n+q+1)},
\end{align}
with $n=\{S, S-1, \cdots, -S+|q|\}$.
Comparing Eqs.($\rm\ref{16}$)-($\rm\ref{18}$) and Eqs.($\rm\ref{25}$)-($\rm\ref{27}$), these matrices are equivalent to those in the model ($\rm\ref{Sz1}$) except for a constant multiplication and a sum of scalar multiplication $-|\gamma_{\mu}|^{2}q^{2}/S$ of the identity matrix. Therefore, these eigenvalues properties are the same as those in the model ($\rm\ref{Sz1}$) for $\sum_{\mu}|\alpha_{\mu}|^{2}$, $\sum_{\mu}|\beta_{\mu}|^{2}$, $\sum_{\mu}|\gamma_{\mu}|^{2}q^{2}\ll S$. Also, it can be found that the case for $\sum_{\mu}|\alpha_{\mu}|^{2}=\sum_{\mu}|\beta_{\mu}|^{2}$ corresponds to the case for $p=0$ in the model ($\rm\ref{Sz1}$). On the other hand, the case for $\sum_{\mu}|\alpha_{\mu}|^{2}\neq\sum_{\mu}|\beta_{\mu}|^{2}$ corresponds to the case for $p\neq0$ in the model ($\rm\ref{Sz1}$). Therefore, for $\sum_{\mu}|\alpha_{\mu}|^{2}=\sum_{\mu}|\beta_{\mu}|^{2}$, the perturbative analysis up to the first-order corrections shows that the real parts of many eigenvalues approach to 0 and the commensurability holds. This means that the BTCs appear in the first-order corrections. On the other hand, for $\sum_{\mu}|\alpha_{\mu}|^{2}\neq\sum_{\mu}|\beta_{\mu}|^{2}$, the Liouvillian gap is not closed even when $S\to\infty$, and thus the BTC does not appear.

Lastly, we show that the condition $\sum_{\mu}|\alpha_{\mu}|^{2}=\sum_{\mu}|\beta_{\mu}|^{2}$ holds if the model ($\rm\ref{Sx1}$) satisfies the Liouvillian symmetry ($\rm\ref{HuberPT}$) when we choose the parity operator to be the reflection of the basis $S_{z}$. 
From the Liouvillian symmetry ($\rm\ref{HuberPT}$), there exists the Lindblad operator $L_{\mu^{\prime}}$ corresponding $L_{\mu}$,
\begin{align}
\label{LPTmu}
L_{\mu^{\prime}}=\alpha_{\mu^{\prime}}S_{x}^{+}+\beta_{\mu^{\prime}}S_{x}^{-}+\gamma_{\mu^{\prime}}S_{x}=e^{i\theta_{\mu}}(-\beta_{\mu}^{*}S_{x}^{+}-\alpha_{\mu}^{*}S_{x}^{-}+\gamma_{\mu}^{*}S_{x}),
\end{align}
where the asterisk means the complex conjugate and $\theta_{\mu}\in\mathbb{R}$. Here, the arbitrariness of the phase $\theta_{\mu}$ causes from the relation,
\begin{align}
\hat{\mathcal{L}}[H; L_\mu, \mu=1,2,\cdots]=\hat{\mathcal{L}}[H; e^{i\theta_{\mu}}L_\mu,\mu=1,2,\cdots].
\end{align}
From Eqs.($\rm\ref{Lmu}$), ($\rm\ref{LPTmu}$), it holds that 
\begin{align}
\sum_{\mu}|\alpha_{\mu}|^{2}+\sum_{\mu^{\prime}}|\alpha_{\mu^{\prime}}|^{2}=\sum_{\mu}|\alpha_{\mu}|^{2}+|\beta_{\mu}|^{2}=\sum_{\mu^{\prime}}|\beta_{\mu^{\prime}}|^{2}+\sum_{\mu}|\beta_{\mu}|^{2},
\end{align}
and thus, the total gain and loss are exactly balanced.

\begin{figure}[htbp]
   \vspace*{-1.8cm}
     \hspace*{-9.5cm}
\includegraphics[bb=0mm 0mm 90mm 150mm,width=0.24\linewidth]{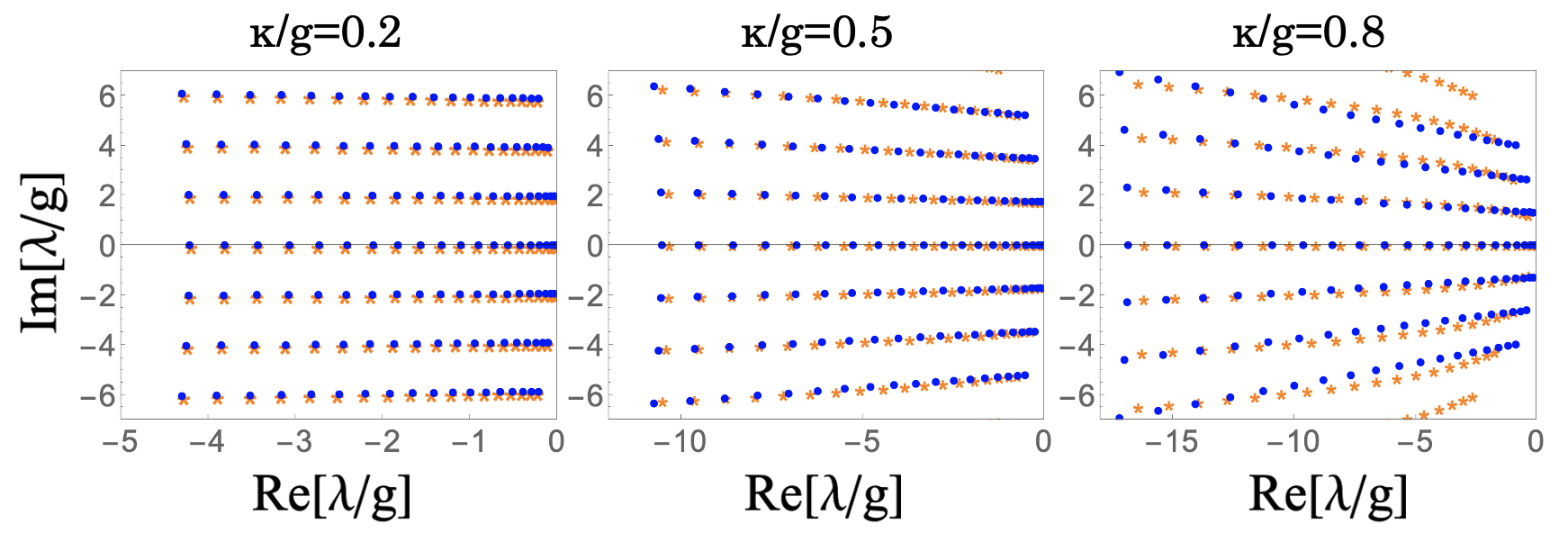}
    \caption{Comparison between the numerical Liouvillian diagonalization (orange star) and the perturbation analysis result up to the second-order corrections (blue circle) for the one-spin BTC model for $\kappa=0.2, 0.5, 0.8$ when $S=10$.}
    \label{fig1spinBTCper}
\end{figure}

\subsection{Perturbation theory for the one-spin BTC model}
Lastly, we apply the degenerate perturbation theory to the one-spin BTC model. We choose that the non-perturbative part is a coherent part in Eq.($\rm\ref{1spinBTC}$), namely $\hat{\mathcal{L}}_{0}[\cdot]=-i[H,\cdot]$, and the perturbative part is the dissipation part in Eq.($\rm\ref{1spinBTC}$). Since this model is equivalent to the model Eq.($\rm\ref{Sx1}$) for $L_{1}=(S_{x}-i(S_{x}^{+}+S_{x}^{-})/2)/\sqrt{2}$ except for a constant multiplication and it satisfies Liouvillian $\mathcal{PT}$ symmetry ($\rm\ref{HuberPT}$), the BTC appears in the first-order corrections.

Now, we consider the second-order correction. The second-order eigenvalue corrections are always pure imaginary numbers since the numerator, and denominator of the fraction in Eq.($\rm\ref{2dp}$) are real and pure imaginary, respectively. Following the prescription of the degenerate perturbation theory, the second-order eigenvalue corrections can be numerically calculated as in Fig.$\rm\ref{fig1spinBTCper}$. 

\begin{figure}[htbp]
   \vspace*{1cm}
     \hspace*{-5.9cm}
\includegraphics[bb=0mm 0mm 90mm 150mm,width=0.23\linewidth]{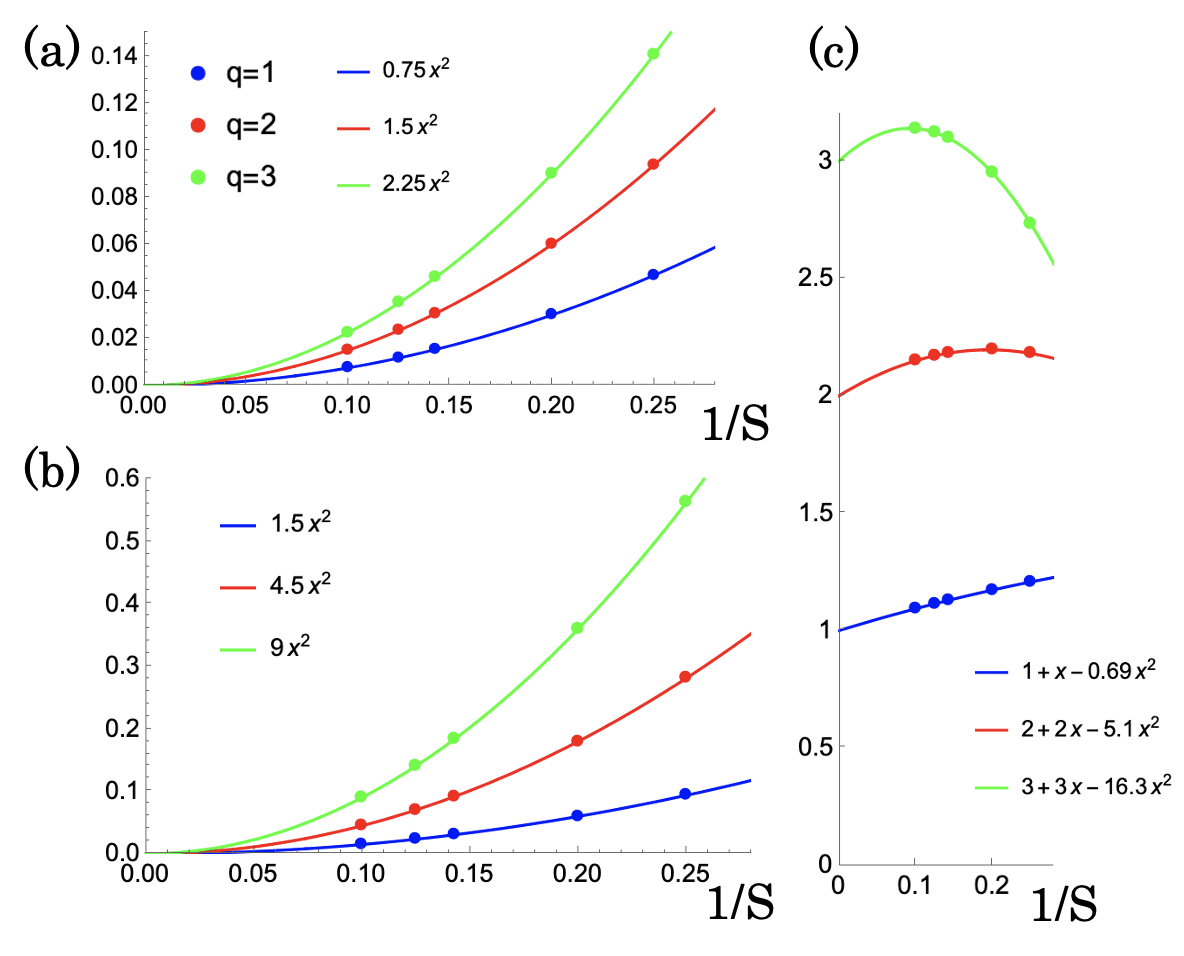}
    \caption{(a) The 2-difference of the second-order eigenvalue correction for $1/S$. (b) The difference between the first and second maximum second-order eigenvalue correction for $1/S$. (c) the maximum second-order eigenvalue correction for $1/S$. These results show that the absolute value of the imaginary part of eigenvalues except for $q=0$ decreases to 0 while retaining the commensurability of the imaginary part of eigenvalues for $S\gg1$.}
    \label{fig1spinBTCkai}
\end{figure}

Furthermore, we numerically find that the second-order eigenvalues correction $\tilde{\lambda}_{m,n}^{(2)}$ for each sector $q$ is the 2-Arithmetic progressions. The 2-difference of the second-order eigenvalue correction for $1/S$ is plotted for each sector $q$ in Fig.$\rm\ref{fig1spinBTCkai}$ (a). Fitting with a quadratic function, it can be seen that the 2-differences approach zero when $S\to\infty$.
The $1/S$ dependence of the difference of the first and second maximum second-order eigenvalue correction is plotted for each sector $q$ in Fig.$\rm\ref{fig1spinBTCkai}$ (b). Fitting with a quadratic function, it can be seen that they approach zero when $S\to\infty$. 
The $1/S$ dependence of the maximum second-order eigenvalue correction is plotted for each sector $q$ in Fig.$\rm\ref{fig1spinBTCkai}$ (c). Fitting with a quadratic function, it can be seen that they approach each sector $q$ when $S\to\infty$.  These results show that the absolute value of the imaginary part of eigenvalues except for $q=0$ decreases to 0 while retaining the commensurability of the imaginary part of eigenvalues for $S\gg1$. In other words, we can catch the behavior of the BTC phase transition up to the second-order correction.

\end{document}